\documentclass[12pt,a4paper,english]{JHEP3}
\usepackage[T1]{fontenc}
\usepackage[latin1]{inputenc}
\usepackage{graphicx}

\makeatletter

\usepackage{graphicx}

\usepackage{amsmath,cite,graphicx}
\title{{\textbf{A Simulation of QCD Radiation in Top Quark Decays.}}}
\author{Keith Hamilton and Peter Richardson \\ Institute of Particle Physics Phenomenology, Department of Physics \\ University of Durham,  Durham, DH1 3LE, UK \\ Email: \email{keith.hamilton@durham.ac.uk, peter.richardson@durham.ac.uk}} \preprint{IPPP/06/96 \\DCPT/06/192}
\abstract{In this paper we describe a theoretical framework and algorithms for implementing QCD
 corrections to top quark decays in the {\sf{Herwig++}} event generator. The dominant corrections, due to soft and collinear emissions, are summed to all orders through the coherent parton branching formalism. In addition, unenhanced first-order matrix-element corrections are included to account for large transverse momentum emissions.}

\newcommand{\newc}{\newcommand}
\newc{\gev}{\,GeV}
\newc{\mev}{\,MeV}
\newc{\ra}{\rightarrow}
\newc{\rpv}{$\mathrm{\not\!R_p}$}
\newc{\rp}{$\mathrm{R_p}$}
\newc{\real}{\mathcal{R}e}
\newc{\alsm}{{\displaystyle \sum_{\alpha=1,2}}}
\newc{\besm}{{\displaystyle \sum_{\beta=1,2}}}
\newc{\al}{\alpha}
\newc{\sgn}{\mr{sgn}\,}
\newc{\be}{\beta}
\newc{\ga}{\gamma}
\newc{\de}{\delta}
\newc{\sla}{\!\!\!\!\!\not\:\:\!}
\newc{\slab}{\!\!\!\!\!\not\,\,\,}
\newc{\slac}{\!\!\!\!\!\!\!\not\,\,\,\,}
\newc{\met}{$\not\!\!E_T$}
\newc{\cw}{\cos\theta_W}
\newc{\sw}{\sin\theta_W}
\newc{\ssw}{\sin^2\theta_W}
\newc{\ccw}{\cos^2\theta_W}
\newc{\cbe}{\cos\beta}
\newc{\sbe}{\sin\beta}
\newc{\ort}{\frac1{\sqrt{2}}}
\newc{\sh}{\hat{s}}
\newc{\uh}{\hat{u}}
\newc{\tha}{\hat{t}}
\newc{\sa}{\sin\al}
\newc{\ca}{\cos\al}
\newc{\mz}{M_{\mr{Z}}}
\newc{\mw}{M_{\mr{W}}}
\newc{\bv}{$\mathrm{\not\!B}$}
\newc{\lv}{$\mathrm{\not\!L}$}
\newc{\beq}{\begin{equation}}
\newc{\eeq}{\end{equation}}
\newc{\ie}{{\it i.e.\/}\ }
\newc{\lam}{\lambda}
\newc{\cht}{\tilde{\chi}}
\newc{\glt}{\tilde{g}}
\newc{\upt}{\tilde{u}}
\newc{\qkt}{\tilde{q}}
\newc{\elt}{\tilde{\ell}}
\newc{\hgt}{\tilde{H}}
\newc{\nut}{\tilde{\nu}}
\newc{\dnt}{\tilde{d}}
\newc{\ftl}{\mr{\tilde{f}}}
\newc{\psb}{\bar{\psi}}
\newc{\rtt}{\sqrt{2}}
\newc{\mut}{\tilde{\mu}}
\newc{\mr}{\mathrm}
\newc{\bath}{\bar{\theta}}
\newc{\tht}{\theta}
\newc{\JC}{{\bf J}}
\newc{\lra}{\longrightarrow}
\newc{\eg}{{\it e.g.\  }}
\newc{\barr}{\begin{eqnarray}}
\newc{\earr}{\end{eqnarray}}
\newc{\me}{\mathcal{M}}
\newc{\dbm}{\partial_\mu}
\newc{\dbmu}{\stackrel{\leftrightarrow\  }{\partial^\mu}}
\newc{\sgm}{\sigma_\mu}
\newc{\captionB}[2]{\caption[{#1}]{{\small {#2}}}}
\usepackage{babel}

\usepackage{babel}
\makeatother
\begin{document}

\section{Introduction}

In the coming era particle physics experiments will probe ever greater
energies so naturally top quark physics will play an increasingly
important r\^{o}le in experimental and theoretical studies. A major
area of study at forthcoming collider experiments will be the precision
measurement of the top quark mass, which is one of the fundamental
input parameters of the standard model and gives rise to the leading
contributions to its effective potential. In addition, a thorough
understanding of top quark physics is crucial for the discovery of
new heavy particles, most notably the Higgs boson, since, due their
copious rate of production, top quarks will provide the main source
of standard model background.

Although inclusive quantities such as total cross sections are well
described by fixed-order QCD calculations, experimental analyses require
a detailed description of the final state. Large logarithmic contributions
to differential distributions must be resummed to all orders and hadronization
effects taken into account. Parton shower Monte Carlo simulations
include these higher-order corrections by appealing to the strongly
ordered soft/collinear limit in which higher-order matrix elements
are represented by universal factors multiplying the lowest order
matrix element. This leading-log approximation scheme may be recast
in a probabilistic form, a Markov chain, from which we can attempt
to generate events as they occur in nature. 

In this paper we will describe such a prescription for the simulation
top quark decays, implemented in the new \textsf{Herwig++} event generator
\cite{Gieseke:2006ga,Gieseke:2006rr}. Many issues raised here and
aspects of the physics we present are also relevant to a discussion
of radiation from the top quark in its production. In full generality
it is not possible to study the production and decay phases separately,
doing so corresponds to working in the \emph{narrow} \emph{width}
\emph{approximation}, \emph{i.e.} neglecting the width of the top
quark. In practice this means the width of the top quark, which is
of order 1 GeV, should be considered infinitesimal and so we should
not simulate gluons with energies below that scale. This does not
present a problem as such a scale is already approximately equal to
the typical cut-off scale used to terminate the parton shower (the
scale at which the parton shower hands over to the hadronization model). 

Simulations of this process have been considered in the past, in the
older \texttt{}\textsf{FORTRAN} HERWIG program \cite{Corcella:1998rs,Corcella:2002jc}.
The simulation which we describe improves on this earlier work in
a number of ways, partially due to general theoretical developments
\cite{Gieseke:2003rz,Gieseke:2003hm}. The older HERWIG \textsf{}program
was based on a non-covariant showering formalism. This meant that,
working in the top quark rest frame, the top quark could emit no radiation
before it decayed. Consequently the entire phase-space had to be populated
as though all emissions originated from the \emph{b}-quark. In addition,
the older program had grown out of a formalism based on massless emitting
partons which forced the introduction of an \emph{ad hoc} angular
cut-off on gluon emission, giving rise to an unphysical \emph{halo}
of gluons at small angles \cite{Marchesini:1989yk}. 

In this paper we describe an approach based on a new covariant \textsf{}parton
showering formalism, which naturally includes gluon radiation from
the decaying top quark \cite{Gieseke:2003rz}. This has the advantage
that the region of phase space corresponding to soft gluon emission
is populated exclusively by the parton shower, instead of the single
gluon emission matrix element, as in earlier simulations, which required
an \emph{ad} \emph{hoc} soft \emph{}cut-off \cite{Corcella:1998rs,Corcella:2002jc}.
Another significant benefit of this new formalism is that it enables
a correct treatment of the masses of the emitting partons through
the use of quasi-collinear splitting functions: no angular cut-off
is required. 

Finally we note that the standard coherent parton shower algorithm
has two important drawbacks. Firstly, because the parton shower generates
emissions from each leg of the hard scattering (quasi-) \emph{independently},
each additional emission must be \emph{uniquely} associated to a particular
leg of the hard scattering, which can only be achieved at the price
of having regions of phase space, corresponding to high $p_{T}$ gluon
emissions, which are unpopulated by the shower. Secondly, the soft/collinear
approximation to the QCD matrix elements is not a good approximation
all over the phase-space region populated by the parton shower. Both
of these problems may be solved by so-called \emph{matrix element
corrections} \cite{Seymour:1994df} which ensure that the hardest
additional gluon emission in the event is distributed according to
the \emph{exact} matrix element. We discuss the inclusion of these
corrections in the approach of \cite{Gieseke:2003rz} which leads
to significant improvements in predictions of physical observables.

In the next section we present the basic parton shower formalism,
based on the covariant parton shower formalism described in \cite{Gieseke:2003rz}.
As we are using the parton shower approximation in a range of kinematics
and masses that it has not been used in before%
\footnote{In the earlier work of \cite{Corcella:1998rs} only the \emph{b}-quark
radiated.%
} our discussion is accordingly detailed. The inclusion of the matrix
element correction to the decays is considered in section~\ref{sec:Matrix-Element-Corrections}
followed by a discussion of the results of the simulation in section~\ref{sec:Results}.
Finally we present our conclusions and plans for further developments.

\section{Basic shower formulation for \emph{t}$\rightarrow$\emph{bW} decays\label{sec:Showering}}

Unlike other quarks, the mass and width of the top quark are such
that it emits radiation and decays before hadronization occurs \cite{Bigi:1986jk,Khoze:1992rq}.
In this section we describe the parton shower approximation to the
decay of a top quark to a $W$ boson and a $b$-quark with additional
gluon radiation. Here we are only concerned with the \emph{decay}
of the top quark, that is our initial state is an \emph{on-shell}
top quark produced in some hard scattering process.

\subsection{Shower variables\label{sub:Shower-variables.} }

\begin{figure}[t]
\begin{center}\includegraphics[%
  width=0.70\textwidth,
  keepaspectratio]{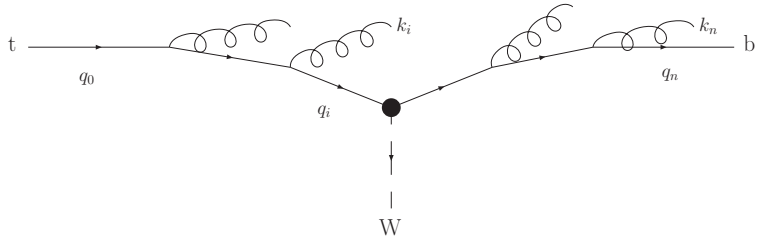}\end{center}

\caption{\label{fig:top_branching}An example of the decay $t\rightarrow bW+\left(n\right)g$.}
\end{figure}

We use the conventions and shower variables described in \cite{Gieseke:2003rz},
which we briefly review here; these variables, together with an appropriate
ordering condition for multiple emissions, ensure that the emissions
obey the angular ordering of QCD radiation. 

The decay $t\rightarrow bW+\left(n\right)g$ is depicted in figure
\ref{fig:top_branching}. The parton showers may be viewed as cascades
of quarks decaying to quark-gluon pairs: $q_{i-1}\rightarrow q_{i}+k_{i}$.
The momenta of the quarks and gluons in the shower are defined in
terms of the following Sudakov decomposition\begin{equation}
\begin{array}{lcl}
q_{i} & = & \alpha_{i}p+\beta_{i}n+q_{\perp i},\\
k_{i} & = & q_{i-1}-q_{i}.\end{array}\label{eq:2.1.1}\end{equation}
As usual in parton shower simulations, prior to simulating gluon emissions
the leading order decay must be generated. The initial (on-shell)
top and bottom quark momenta serve to define the basis vectors $p$
needed for each shower. The $n$ and $q_{\perp i}$ basis vectors
are defined in the rest frame of the initial top quark with the initial
\emph{b}-quark momentum in the $z$ direction. In this frame the $n$
vectors for top and \emph{b}-quark showers are chosen to be\begin{equation}
\begin{array}{lcl}
n_{t} & = & \frac{1}{2}m_{t}\left(1,\mathbf{0},1\right),\\
n_{b} & = & \frac{1}{2}m_{t}\left(\lambda,\mathbf{0},-\lambda\right),\end{array}\label{eq:2.1.2}\end{equation}
where $\lambda$ is given by,\begin{equation}
\lambda=\frac{1}{m_{t}^{2}}\sqrt{m_{t}^{4}+m_{W}^{4}+m_{b}^{4}-2m_{t}^{2}m_{b}^{2}-2m_{t}^{2}m_{W}^{2}-2m_{W}^{2}m_{b}^{2}}.\label{eq:2.1.2.b}\end{equation}
 The $q_{\perp i}$ vectors have the form $\left(0,\mathbf{q}_{\perp i},0\right)$
in this frame. 

In addition to the Sudakov decomposition (\ref{eq:2.1.1}) we also
adopt the following variables from \cite{Gieseke:2003rz} for the
top quark shower\begin{equation}
\begin{array}{lclclcl}
\tilde{q}_{i}^{2} & = & \frac{m_{t}^{2}-q_{i}^{2}}{1-z_{i}}, & \,\,\,\,\,\,\,\, & z_{i} & = & \frac{\alpha_{i}}{\alpha_{i-1}},\\
p_{\perp i} & = & q_{\perp i}-z_{i}q_{\perp i-1}, & \,\,\,\,\,\,\,\, & \mathbf{p}_{\perp i}^{2} & = & -p_{\perp i}^{2},\end{array}\label{eq:2.1.3}\end{equation}
where $p_{\perp i}$ is defined to be the relative transverse momentum
involved in branching $i$, and $\tilde{q}_{i}^{2}$ is the \emph{evolution}
\emph{variable}. For radiation from the bottom quark the definitions
in (\ref{eq:2.1.3}) are unchanged except for the evolution variable,
which becomes \begin{equation}
\tilde{q}_{i}^{2}=\frac{q_{i-1}^{2}-m_{b}^{2}}{z_{i}\left(1-z_{i}\right)}.\label{eq:2.1.4}\end{equation}

These evolution variables $\tilde{q}_{i}^{2}$ are defined by close
analogy to those of the original HERWIG program \cite{Corcella:2002jc},
they correspond closely to the angle between the emitted quark and
gluon: ordering in $\tilde{q}_{i}^{2}$ results in angular ordering
of both the top quark and bottom quark parton showers. This feature
is discussed in more detail in section (\ref{sub:Soft-gluon-coherence}).

\subsection{Shower phase space}

We begin by considering the decay $t\rightarrow bW+\left(n\right)g$,
where the gluons are \emph{unresolved}. The partial width for such
a decay is given by\begin{equation}
\int\mathrm{d}\Gamma_{n}=\frac{1}{2m_{t}}\int\mathrm{d}\Phi_{bW}\mathrm{d}\Phi_{K}\textrm{ }\left(2\pi\right)^{4}\delta^{4}\left(p_{t}-q_{b}-q_{W}-K\right)\left|\mathcal{M}_{n}\right|^{2},\label{eq:2.2.1}\end{equation}
where $m_{t}$ is the mass of the top, $p_{t}$ is its four-momentum
and $q_{b/W}$ is the four-momentum of the $b/W$. We have also denoted
the matrix element $\mathcal{M}_{n}$. $K$ denotes the sum of the
individual gluon momenta $\sum_{i=1}^{n}k_{i}$, while $\mathrm{d}\Phi_{K}$
and $\mathrm{d}\Phi_{bW}$ are the phase-space measures for the gluon
and $bW$ systems respectively,\begin{equation}
\begin{array}{lcl}
\int\mathrm{d}\Phi_{K} & = & \prod_{i=1}^{n}\int\mathrm{d}\Phi_{k_{i}}=\prod_{i=1}^{n}\int_{\mathcal{U}}\frac{\mathrm{d}^{3}k_{i}}{\left(2\pi\right)^{3}2k_{i}^{0}},\\
\int\mathrm{d}\Phi_{bW} & = & \prod_{i=b,W}^{n}\int\frac{\mathrm{d}^{3}q_{i}}{\left(2\pi\right)^{3}2q_{i}^{0}}.\end{array}\label{eq:2.2.2}\end{equation}
 The symbol $\mathcal{U}$ denotes the region of phase space \emph{inside}
which gluons are unresolvable.

Assuming the branching picture depicted in figure \ref{fig:top_branching},
the full phase space may be factorised by repeatedly inserting the
identity as integrals over two simple delta functions, one such insertion
for each gluon vertex:\begin{equation}
\begin{array}{rcl}
\int\mathrm{d}\Phi_{bW}\mathrm{d}\Phi_{K} & = & \prod_{i=1}^{n}\int\mathrm{d}\Phi_{i}\int\mathrm{d}\Phi_{bW}\left(2\pi\right)^{4}\delta^{4}\left(q_{i}-q_{i+1}-q_{W}\right)\\
\int\mathrm{d}\Phi_{i} & = & \int\mathrm{d}\Phi_{k_{i}}\mathrm{d}^{4}q_{i}\mathrm{d}Q_{i}^{2}\textrm{ }\delta\left(Q_{i}^{2}-q_{i}^{2}\right)\delta^{4}\left(q_{i-1}-q_{i}-k_{i}\right)\end{array},\label{eq:2.2.3}\end{equation}
where $q_{0}=p_{t}$, $q_{n}=q_{b}$. Given the branching picture
in figure \ref{fig:top_branching}, the parton showers may be viewed
as cascades of two body decays $q\rightarrow qg$. This interpretation
is evident from the fact that $\mathrm{d}\Phi_{i}$ has the familiar
form of a two body phase-space measure for a quark of mass $Q_{i}$
and a gluon, albeit with an additional $Q_{i}^{2}$ integration.

Exploiting the Lorentz invariance of the integration measure, we may
rewrite the two body phase-space integrals for the top quark (\ref{eq:2.2.3})
as \begin{equation}
\int\mathrm{d}\Phi_{i}=\frac{1}{4\left(2\pi\right)^{2}}\int\mathrm{d}\tilde{q}_{i}^{2}\mathrm{d}z_{i}\textrm{ }\left(1-z_{i}\right),\label{eq:2.2.4}\end{equation}
these differ from those of the \emph{b}-quark by a factor of $z_{i}$
due to slightly different definitions of the evolution variables (\ref{eq:2.1.3},\ref{eq:2.1.4}).

\subsection{Soft gluon coherence and angular ordering\label{sub:Soft-gluon-coherence}}

The integration limits on $\tilde{q}_{i}^{2}$ and $z_{i}$ in (\ref{eq:2.2.4})
may be inferred from the kinematic constraints involved in each splitting,
however, as we shall now discuss, \emph{dynamical}, soft gluon interference,
effects motivate further restrictions on this phase space and hence
a modification to the integration bounds implied by the kinematics.

Using the soft gluon insertion technique \cite{Bassetto:1984ik},
and following a similar analysis to that in \cite{Ellis:1991qj},
the squared amplitude for a soft gluon dressing the decay $t\rightarrow bWg$,
with amplitude $\mathcal{M}_{1}$, is given by\begin{eqnarray}
\lim_{k\rightarrow0}\left|\mathcal{M}_{2}\right|^{2} & = & \frac{2g_{s}^{2}}{\omega^{2}}\left(\begin{array}{ll}
 & C_{F}\left(W_{tg}^{t}+\tilde{W}_{gb}^{t}-\tilde{W}_{tb}^{g}+\tilde{W}_{gt}^{b}\right)\\
+ & C_{F}\left(\tilde{W}_{gb}^{t}+\tilde{W}_{tb}^{g}+\frac{1}{2}W_{tb}^{b}+\frac{1}{2}W_{bg}^{b}\right)\\
+ & C_{A}\left(W_{tg}^{g}+\tilde{W}_{tb}^{g}-\tilde{W}_{gb}^{t}+\tilde{W}_{tg}^{b}\right)\end{array}\right)\left|\mathcal{M}_{1}\right|^{2}\label{eq:2.3.1}\end{eqnarray}
where\begin{equation}
\begin{array}{rcl}
W_{ij}^{i} & = & \frac{1}{2n_{i}.n_{k}}\left(1-\frac{n_{i}^{2}}{n_{i}.n_{k}}+\frac{n_{i}.n_{j}-n_{i}.n_{k}}{n_{j}.n_{k}}\right),\\
\tilde{W}_{jk}^{i} & = & \frac{1}{2}\left(W_{ik}^{i}-W_{ij}^{i}\right),\end{array}\label{eq:2.3.2}\end{equation}
$\omega$ is the energy of the soft gluon, $n_{i}=p_{i}/E_{i}$, $n_{k}=k/\omega$,
and the sum over colours and spins, is understood. The first two terms
proportional to $C_{F}$ are due to (incoherent) radiation from the
top quark and bottom quark respectively, the third term, proportional
to $C_{A}$, is due to radiation from the hard gluon $\left(g\right)$.
Averaging $W_{ij}^{i}$ over the azimuthal angle, about $p_{i}$,
one readily finds \begin{equation}
\left\langle W_{ij}^{i}\right\rangle =\frac{1}{2n_{i}.n_{k}}\left(1-\frac{n_{i}^{2}}{n_{i}.n_{k}}+\frac{v_{i}\left(n_{i}.n_{j}-n_{i}.n_{k}\right)}{\sqrt{\left(n_{i}.n_{j}-n_{i}.n_{k}\right)^{2}-n_{j}^{2}\left(\left(1-n_{i}.n_{k}\right)^{2}-v_{i}^{2}\right)}}\right),\label{eq:2.3.3}\end{equation}
where $n_{i}^{2}=1-v_{i}^{2}$, with $v_{i}$ the velocity of particle
$i$. Clearly in the limit $v_{i},v_{j}\rightarrow1$, we see that
the emission from $i$ in a pair of partons $ij$ is restricted to
a cone \emph{viz}\begin{equation}
\left\langle W_{ij}^{i}\right\rangle =\frac{1}{2n_{i}.n_{k}}\theta\left(\theta_{ij}-\theta_{i}\right),\label{eq:2.3.4}\end{equation}
where $\theta_{i}$ is the angle between the soft gluon and $i$,
and $\theta_{ij}$ is the angle between $i$ and $j$. The vanishing
of the radiation outside the cone is attributable to destructive interference.

To demonstrate the angular ordering of the radiation from the top
quark we need to consider the limit in which the directions of the
top quark and hard gluon approach one another. We also need to consider
$v_{t},v_{b}\rightarrow1$ in order to use (\ref{eq:2.3.4}). In the
dipole rest frame (the rest frame of the $W$ boson) $v_{t}$ is around
0.7, so the validity of working in the limit $v_{t},v_{b}\rightarrow1$
is questionable, we will address this matter shortly. Proceeding with
these approximations we find $n_{g}=n_{t}$ and hence\begin{equation}
\begin{array}{rcl}
\left|\mathcal{M}_{2}\right|^{2} & = & \frac{2g_{s}^{2}}{\omega^{2}}\left(C_{F}W_{tg}^{t}+C_{A}W_{tg}^{g}+C_{F}W_{t^{*}b}^{b}+C_{F}\tilde{W}_{t^{*}b}^{t^{*}}\right)\left|\mathcal{M}_{1}\right|^{2}\\
W_{t^{*}b}^{b} & = & \frac{1}{2}\left(W_{tb}^{b}+W_{bg}^{b}\right)\\
\tilde{W}_{t^{*}b}^{t^{*}} & = & \frac{1}{2}\left(W_{tb}^{t}-W_{tg}^{t}\right)+\frac{1}{2}\left(W_{gb}^{g}-W_{gt}^{g}\right)\end{array}.\label{eq:2.3.5}\end{equation}

The first two terms in the radiation pattern (\ref{eq:2.3.5}) are
due to (incoherent) emissions from the top quark and hard gluon respectively.
The third term in the radiation pattern is due to (incoherent) radiation
from the \emph{b}-quark, due to its interaction with $t^{*}$, by
which we mean the amalgamation of $t$ and $g$ (soft gluon emissions
from $b$ will not `resolve' the separation of $t$ and $g$). In
the limit $v_{t},v_{b}\rightarrow1$, the azimuthal average of each
of the first three terms, of the form $W_{ij}^{i}$, is given by (\ref{eq:2.3.4}),
so soft gluon emissions are restricted to cones whose half angle is
equal to that between the two particles forming the colour dipole
($i$ and $j$).

Similarly, by virtue of (\ref{eq:2.3.4}), the radiation due to the
fourth term in (\ref{eq:2.3.5}) is restricted to a \emph{wide} cone
lying along the $t^{*}$ direction, reaching out to the \emph{b}-quark,
but with emissions \emph{vetoed} in the smaller cones, where the incoherent
$t$ and $g$ emissions are allowed. The radiation due to the fourth
term is considered to be the coherent sum of emissions from $t$ and
$g$ (\ref{eq:2.3.5}) and therefore, equivalently, soft, angular
ordered, wide angle, radiation from the internal, off-shell $t^{*}$
line. \emph{}This analysis indicates that in evolving toward the decay,
from the on-shell top quark, the angle of the emissions is always
increasing. \emph{}

We now turn to question the validity of using the $v_{t},v_{b}\rightarrow1$
limit in motivating this angular ordering. In \cite{Marchesini:1989yk}
it was found that \emph{equal} finite parton masses in $\left\langle W_{ij}^{i}\right\rangle $
smooth the angular cut-off provided by the step function in (\ref{eq:2.3.4})
to the form shown in brackets in (\ref{eq:2.3.3}), also the finite
masses screen the collinear singularity in the $2n_{i}.n_{k}^{-1}$
prefactor giving rise to the so-called \emph{dead-cone} effect \cite{Marchesini:1989yk}. 

\begin{figure}[t]
\begin{center}\includegraphics[%
  clip,
  width=0.50\textwidth,
  keepaspectratio]{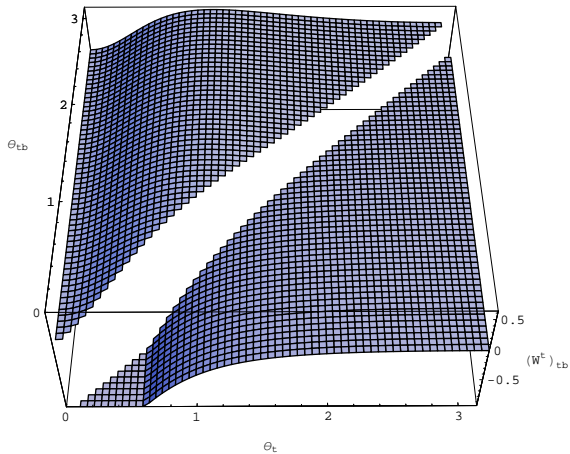}\end{center}

\caption{\label{fig:w_ij}In this figure we show the azimuthally averaged
radiation pattern $\left\langle W_{tb}^{t}\right\rangle $, for $v_{t}=0.65$,
$v_{b}=1$. In this plot, the ratio of the integral of $\left\langle W_{tb}^{t}\right\rangle $
outside the cone, to the same integral inside the cone is around $-0.33$. }
\end{figure}

In the case of the $t\rightarrow bW$ decay, assigning velocities
to $t$ and $b$ like those in the $W$ rest frame, the dead\emph{-}cone
effect is pronounced but the smoothing of the $\theta\left(\theta_{tb}-\theta_{t}\right)$
cut-off does not occur, as one can see from the plot of $\left\langle W_{tb}^{t}\right\rangle $
in figure \ref{fig:w_ij}. On the other hand, in the $t\rightarrow bW$
case, the radiation distribution does not simply vanish outside the
cone like $\theta\left(\theta_{tb}-\theta_{t}\right)$ but rather
it can be negative for small $\theta_{tb}$. Therefore, by performing
conventional angular ordering, we will be making an approximation
\emph{i.e.} neglecting this negative contribution. We argue, as in
\cite{Marchesini:1989yk}, that this is justified given that the majority
of the radiation is inside the cone: for $v_{t}\approx0.7$, $v_{b}\approx1$,
as in figure \ref{fig:w_ij}, we find that the ratio of the integral
of $\left\langle W_{tb}^{t}\right\rangle $ outside the cone to the
same integral inside the cone is approximately $-0.3$. 

The same calculation can be performed for the \emph{b}-quark in the
limit $v_{b}\rightarrow1$ and one finds that the emission angles
are required to be successively smaller as one evolves away from the
decay. 

In terms of the $\tilde{q}_{i}$ evolution variable, in the top quark
rest frame, in the soft limit $\left(1-z=\epsilon\right)$, one may
show that\begin{equation}
\frac{\tilde{q}_{t}^{2}}{m_{t}^{2}}=\frac{2}{1-\cos\theta_{b}},\label{eq:2.3.6}\end{equation}
where $\theta_{b}$ is the angle between the soft gluon and the \emph{b}-quark
(since the top is at rest) \emph{i.e.} starting at $\tilde{q}_{0}^{2}=m_{t}^{2}$
and evolving to higher values monotonically increases the angle/transverse
momentum of the soft gluon emission with respect to the \emph{b}-quark
direction. This is consistent with our analysis of the soft radiation
pattern from the $t\rightarrow bWg$ decay (\ref{eq:2.3.5}), in the
\emph{W} boson rest frame, which is obtained by a boost in the \emph{W}
boson direction, which will preserve the transverse momentum ordering.
Therefore implementing angular ordering in the top quark shower simply
involves bounding $\tilde{q}_{i+1}>\tilde{q}_{i}$. For the \emph{b}-quark
the situation is the same as in the older HERWIG program namely $\tilde{q}_{i+1}<z_{i}\tilde{q}_{i}$.
These bounds give rise to nested $\tilde{q}_{i}^{2}$ phase-space
integrals. Finally we note that this ordering in $\tilde{q}_{i}^{2}$
is more stringent than the $q_{i}^{2}$ ordering implied by the kinematics,
one may easily prove that the $q_{i}^{2}$ ordering is a byproduct
of the $\tilde{q_{i}}$/angular ordering using just the definitions
in section (\ref{sub:Shower-variables.}).

\subsection{Matrix element approximations\label{sub:Matrix-element-approximations.}}

Thus far we have shown how the phase space for the parton shower may
be exactly factorized in a Lorentz invariant way. We will now briefly
discuss the factorization of the \emph{n}-parton matrix element. 

The principle requirement for factorization of the matrix element
is that the emissions are either soft or quasi-collinear or both.
Such emissions are significantly favoured by the underlying dynamics
and so the majority are of this type. We aim to accurately describe
all quasi-collinear emissions and to take into account soft gluon
emissions through the approximation of angular ordering.

\subsubsection{Emissions from top quarks}

The quasi-collinear limit \cite{Catani:2000ef} is defined to be the
limit in which $\mathbf{p}_{\perp i}^{2}$ and the on-shell quark
mass squared $\left(m_{q}^{2}\right)$ are assumed small compared
to $n.p$ but not compared to each other. For an \emph{n}-parton process
in which a decaying top quark emits a quasi-collinear gluon, the squared
matrix element factorizes according to\begin{eqnarray}
\lim_{q_{n}\parallel k_{n}}\left|\mathcal{M}_{n}\right|^{2} & = & \frac{8\pi\alpha_{\mathrm{S}}}{\left(1-z_{n}\right)\tilde{q}_{n}^{2}}P_{qq}\left(z_{n},\tilde{q}_{n}^{2}\right)\left|\mathcal{M}_{n-1}\right|^{2},\label{eq:2.4.1a}\\
P_{qq}\left(z,\tilde{q}^{2}\right) & = & C_{F}\left(\frac{1+z^{2}}{1-z}-\frac{2zm_{t}^{2}}{\left(1-z\right)\tilde{q}^{2}}\right),\label{eq:2.4.1b}\end{eqnarray}
where $\mathcal{M}_{n-1}$ is the matrix element for the process without
the additional gluon emission. $P_{qq}$ is the quasi-collinear quark-gluon
splitting function \cite{Catani:2000ef}.

Another important requirement for our factorization of the matrix
element is that the gluon emissions are such that the intermediate
quark virtualities are strongly ordered. As mentioned previously,
the virtualities are naturally ordered by kinematics alone. Furthermore,
from the Sudakov decomposition (\ref{eq:2.1.1}) and our definitions
(\ref{eq:2.1.2}), one can show that imposing angular ordering automatically
provides an ordering of virtualities which is more restrictive than
that dictated by the branching picture in figure \ref{fig:top_branching}.
For strongly ordered emissions the decomposition (\ref{eq:2.4.1a})
may be applied recursively, reducing $\left|\mathcal{M}_{n}\right|^{2}$
to a product of splitting functions that multiply the leading-order
matrix element.

Unfortunately, for emissions from the top quark, given our choice
of $n$ (\ref{eq:2.1.2}), $n.p=\frac{1}{2}m_{t}^{2}$. Therefore
one may not expect that the quasi-collinear splitting function approximates
the single emission matrix element very well. Although the choice
of $n$ is arbitrary, in the top quark rest frame $n.p$ only depends
on the energy components of $n$ and, as we shall see shortly (\ref{eq:2.4.2a}),
the quasi-collinear splitting function is invariant under rescalings
of $n$. This is may be viewed as a manifestation of the fact that
collinear enhancements usually take the form of large \emph{mass}
\emph{singular} logarithms of some scale, characteristic of the leading
order process, divided by the quark mass, whereas in this case the
characteristic scale \emph{is} the quark mass. This is, nevertheless,
not a problem; the fact that the collinear emissions are suppressed
means that only soft emissions are enhanced and, as we shall now demonstrate,
these are well modelled by our approximations. 

From the definition of the quasi-collinear splitting function and
the shower variables $z$ and $\tilde{q}^{2}$, we find that the approximation
to the quasi-collinear limit of \emph{}$\left|\mathcal{M}_{n}\right|^{2}$
can be rewritten\begin{eqnarray}
\lim_{q_{i}\parallel k_{i}}\left|\mathcal{M}_{n}\right|^{2} & = & 8\pi\alpha_{\mathrm{S}}C_{F}\mathcal{D}_{t,n}\left|\mathcal{M}_{n-1}\right|^{2},\label{eq:2.4.2a}\\
\mathcal{D}_{t,n} & = & \frac{1}{2q_{i-1}.k_{i}}\left(\frac{n.k_{i}}{n.q_{i-1}}+\frac{2n.q_{i}}{n.k_{i}}-\frac{n.q_{i}m_{t}^{2}}{\left(n.q_{i-1}\right)\left(q_{i-1}.k_{i}\right)}\right).\label{eq:2.4.2b}\end{eqnarray}
In the soft limit the first term in $\mathcal{D}_{t,n}$ is sub-leading
and may be neglected ($k_{i}$ is only in the numerator), also in
the soft limit $q_{i}=q_{i-1}$, so our approximation to $\left|\mathcal{M}_{n}\right|^{2}$
in the soft limit is \begin{equation}
\mathcal{D}_{t,n}=\frac{n.q_{t}}{\left(q_{t}.k_{i}\right)\left(n.k_{i}\right)}-\frac{m_{t}^{2}}{2\left(q_{t}.k_{i}\right)^{2}}.\label{eq:2.4.3}\end{equation}
Neglecting the \emph{b}-quark mass $n\equiv q_{b}$ and we may recognise
this soft limit $\mathcal{D}_{t,n}$ as having \emph{precisely} the
familiar form of the eikonal dipole radiation function, namely,\begin{equation}
\mathcal{D}_{t,n}=-\frac{1}{2}\left(\frac{q_{b}}{q_{b}.k_{i}}-\frac{q_{t}}{q_{t}.k_{i}}\right)^{2}.\label{eq:2.4.4}\end{equation}
Plainly, the quasi-collinear splitting function, together with the
definition of $n$ and the shower variables $\left(z,\textrm{ }\tilde{q}^{2}\right)$,
reproduces exactly the correct soft limit of $\left|\mathcal{M}_{n}\right|^{2}$
if one neglects the mass of the \emph{b}-quark. 

Taking into account the \emph{b}-quark mass, the differences between
the soft approximation we use (\ref{eq:2.4.2a}) and the exact result
(\ref{eq:2.4.4}) will become significant if there are emissions for
which $q_{b}.k_{i}$ becomes small \emph{i.e.} for emissions near-collinear
to the \emph{b}-quark. However such emissions are generated assuming
that they were emitted by the \emph{b}-quark, \emph{i.e.} they are
produced by the \emph{b}-quark shower and not the top quark shower.

The high level of agreement between our approximation (\ref{eq:2.4.2a})
and the exact soft distribution (\ref{eq:2.4.4}) is understandable
given the size of the ratio $m_{b}/m_{t}$. Despite this nice feature,
the distribution of radiation which is produced is nevertheless based
on an approximation, in particular the distribution of high transverse
momentum emissions is known to be modelled poorly. This matter will
be rectified later through the soft-matrix element correction, to
be discussed in section \ref{sec:Matrix-Element-Corrections}. 

Finally, although, strictly speaking, a higher order effect, earlier
studies, neglecting quark mass effects, have shown that a careful
choice of scale for $\alpha_{\mathrm{S}}$ enables one to include
parts of the next-to-leading-log contributions \cite{Amati:1980ch}.
More specifically, for light quarks, the soft $z\rightarrow1$ limit
of the $P_{qq}$ massless splitting function at the two-loop order
may be obtained by expanding $\alpha_{\mathrm{S}}\left(\mathbf{p}_{\perp i}^{2}=\left(1-z\right)^{2}\tilde{q}^{2}\right)$
about $q^{2}$ to $\mathcal{O}\left(\alpha_{\mathrm{S}}^{2}\right)$
and rescaling $\Lambda_{\overline{\mathrm{MS}}}$ . For massive partons
we assume that the appropriate argument for $\alpha_{\mathrm{S}}$
is again $\left(1-z\right)^{2}\tilde{q}^{2}$, which is equal to $\mathbf{p}_{\perp i}^{2}+\left(1-z_{i}\right)^{2}m_{t}^{2}$.
Although the argument of $\alpha_{\mathrm{S}}$ is no longer the local
transverse momentum and although the quasi-collinear splitting function
contains a term not present in the massless case, in the limit $z\rightarrow1$
both the argument of $\alpha_{\mathrm{S}}$ and the quasi-collinear
splitting function tend to their respective values in the massless
analysis and so we can once again expect to reproduce the higher order
terms.

\subsubsection{Emissions from \emph{b}-quarks}

Gluon emissions from \emph{b}-quarks are distributed according to
the associated quasi-collinear splitting function, this is the same
as (\ref{eq:2.4.1b}) but for the replacements $z_{i}m_{t}^{2}\rightarrow m_{b}^{2}$
and $\tilde{q}_{i}^{2}\rightarrow z\tilde{q}_{i}^{2}$. As in the
case of emissions from the top quark, we may express this factorization
in terms of scalar products by using the Sudakov decomposition: \begin{eqnarray}
\lim_{q_{i}\parallel k_{i}}\left|\mathcal{M}_{n}\right|^{2} & = & 8\pi\alpha_{\mathrm{S}}C_{F}\mathcal{D}_{b,n}\left|\mathcal{M}_{n-1}\right|^{2},\label{eq:2.4.5a}\\
\mathcal{D}_{b,n} & = & \frac{1}{2q_{i}.k_{i}}\left(\frac{n.k_{i}}{n.q_{i-1}}+\frac{2n.q_{i}}{n.k_{i}}-\frac{m_{q}^{2}}{q_{i}.k_{i}}\right).\label{eq:2.4.5b}\end{eqnarray}
This splitting function accurately reproduces the distribution of
gluons emitted at small angles to the quark, however, sizeable errors
may occur if the approximation is used beyond the quasi-collinear
limit.

Evidently emissions close to the $n$ direction are enhanced by the
second term in (\ref{eq:2.4.5b}), proportional to $n.k_{i}^{-1}$.
This enhancement is unphysical; as well as being a basis vector of
the Sudakov decomposition, $n$ is also the (axial) gauge vector used
in the calculation of the quasi-collinear splitting functions. The
manifestation of such divergences indicates a breakdown of the quasi-collinear
approximation, \emph{i.e.} emissions for which $q_{i-1}.k\not\gg n.k_{i}$
are beyond the quasi-collinear limit. 

As was shown for the case of the top quark shower, choosing $n$ collinear
with the colour partner gives a good approximation to the eikonal
limit and assigns a physical origin to the $n.k^{-1}$ divergence.
Unfortunately, for the case in hand we cannot choose $n$ collinear
with the top quark since we are in its rest frame and $n$ is light-like.
Instead we choose $n$ acolinear with $q_{b}$, thereby maximally
separating it from the region of phase space into which the shower
can emit gluons. In spite of this, results show a substantial excess
of high transverse momentum emissions from the \emph{b-}quark, due
to the $n.k_{i}^{-1}$ enhancement. This is later rectified by the
soft matrix element correction procedure described in section \ref{sub:Soft-Matrix-Element-Corrections}. 

Although the soft matrix element correction will compensate any excess
emission, we would prefer that this was not the default \emph{modus
operandi}, since we only implement these process-specific corrections
for cases of special interest. By generalizing the quasi-collinear
splitting function to \begin{equation}
P_{qq}\rightarrow\mathcal{V}_{qq}=P_{qq}-C_{F}\left(\frac{q_{i}.k_{i}}{n.k_{i}}\right)\frac{n^{2}}{n.k_{i}}\label{eq:2.4.6}\end{equation}
and setting $n=q_{t}$, we can also remedy the surplus emissions and
generally improve the shower approximation. This modification produces
the correct soft and collinear limits, it is similar in nature to
a soft matrix element correction procedure \cite{Seymour:1994df}.
Unlike the soft matrix element correction, the generalized quasi-collinear
splitting function $\mathcal{V}_{qq}$ is process independent. In
the most basic sense, our splitting function may be viewed as a simple
merging of the quasi-collinear limit with the \emph{full} eikonal
limit of the colour dipole%
\footnote{In this latter respect $\mathcal{V}_{qq}$ differs from the dipole
splitting functions of \cite{Catani:2002hc}: reproducing the eikonal
limit of the colour dipole requires a sum of \emph{two} dipole splitting
functions.%
}. However, our generalization is also motivated from more formal considerations
\emph{viz} working with a more general class of gauges, which we discuss
in appendix \ref{sec:Generalization_of_Pqq}.

\subsection{Sudakov form factor and shower algorithm}

Proceeding with strong ordering and quasi-collinear factorization
as simplifying assumptions, this corresponds to working in the leading-log
approximation. It is well known that large logarithms associated to
collinear and soft emissions must vanish when the full phase-space
is integrated over, this is a consequence of the Block-Nordsieck and
KLN theorems. We may use this fact to rewrite the integrals over the
phase space for unresolved gluon emissions $\mathcal{U}$ (where we
take $\mathcal{U}$ to be the region $z>z_{c}$ where $z_{c}$ is
some cut-off) in terms of integrals over the remaining resolved gluon
phase-space $\mathcal{R}$, specifically we use,\begin{equation}
\int_{\mathcal{U}}\mathrm{d}P_{i}\left(t\rightarrow tg\right)+\int_{\mathcal{R}}\mathrm{d}P_{i}\left(t\rightarrow tg\right)=0\label{eq:2.5.1}\end{equation}
 where \begin{equation}
\int_{\mathcal{U}/\mathcal{R}}\mathrm{d}P_{i}\left(t\rightarrow tg\right)=\int_{\tilde{q}_{i-1}^{2}}^{\tilde{q}_{max}^{2}}\mathrm{d}\tilde{q}_{i}^{2}\mathrm{d}z_{i}\textrm{ }\frac{\alpha_{\mathrm{S}}}{2\pi\tilde{q}_{i}^{2}}P_{qq}\left(z_{i},\tilde{q}_{i}^{2}\right)\Theta\left(\mathcal{U}/\mathcal{R}\right),\label{eq:2.5.2}\end{equation}
and $\Theta\left(\mathcal{U}/\mathcal{R}\right)=1$ for gluon emissions
in $\mathcal{U}/\mathcal{R}$ and zero otherwise. This gives the following
expression for the decay width, with \emph{n} \emph{unresolved} branchings,\begin{equation}
\int\mathrm{d}\Gamma_{n}=\prod_{i=1}^{n}\left(-\int_{\mathcal{R}}\mathrm{d}P_{i}\left(t\rightarrow tg\right)\right)\textrm{ }\frac{1}{2m_{t}}\int\mathrm{d}\Phi_{bW}\left(2\pi\right)^{4}\delta^{4}\left(q_{t,n}-q_{b}-q_{W}\right)\left|\mathcal{M}_{0}\right|^{2}.\label{eq:2.5.3}\end{equation}
Within leading-log accuracy we may neglect the dependence of the final
decay, to the \emph{b}-quark and \emph{W} boson, on $q_{t,n}$. Such
approximations are a generic feature of all leading-log parton shower
simulations. With this in mind we may simply rewrite the nested integrals
over the branching probabilities as \begin{equation}
\begin{array}{lcl}
\int\mathrm{d}\Gamma_{n} & = & \frac{1}{n!}\left(-\int_{\mathcal{R}}\mathrm{d}P_{1}\left(t\rightarrow tg\right)\right)^{n}\textrm{ }\int\mathrm{d}\Gamma_{0}\end{array}.\label{eq:2.5.4}\end{equation}
Summing over all $n$, the corrections due to unresolved gluon emission
exponentiate to give, in the leading-log approximation, \begin{equation}
\Gamma_{\mathcal{U}}=\exp\left(-\int_{\mathcal{R}}\mathrm{d}P_{1}\left(t\rightarrow tg\right)\right)\int\mathrm{d}\Gamma_{0}.\label{eq:2.5.5}\end{equation}
Since all soft and collinear logarithmic corrections to the width
must vanish on integrating over the entire phase-space $\mathcal{R}+\mathcal{U}$,
in the leading log approximation the total width is simply equal to
that given by the leading-order calculation: $\Gamma_{0}$. Hence,
the probability that the top quark \emph{evolves} from scale $\tilde{q}_{0}^{2}$
to $\tilde{q}_{max}^{2}$ without emitting any resolvable radiation
is, \begin{equation}
\Delta\left(\tilde{q}_{0}^{2},\tilde{q}_{max}^{2}\right)=\frac{\Gamma_{\mathcal{U}}}{\Gamma_{0}}=\exp\left(-\int_{\mathcal{R}}\mathrm{d}P_{1}\left(t\rightarrow tg\right)\right),\label{eq:2.5.6}\end{equation}
 the Sudakov form factor. By the same token $\mathrm{d}P_{i}$ represents
the probability of a branching $t\rightarrow tg$ with $z_{i}$ in
the interval $\left[z_{i},z_{i}+\mathrm{d}z_{i}\right]$ and $\tilde{q}_{i}^{2}$
in $\left[\tilde{q}_{i}^{2},\tilde{q}_{i}^{2}+\mathrm{d}\tilde{q}_{i}^{2}\right]$,
for resolvable gluons%
\footnote{With this interpretation (\ref{eq:2.5.1}) implies that this same
probability $\mathrm{d}P_{i}$ is negative in the unresolvable region.
Clearly the interpretation of $\mathrm{d}P_{i}$ as a probability
in this region is not sensible, however the minus sign is expected,
it is known to originate from the virtual gluon emissions (loop contributions),
which are of course unresolvable emissions.%
}. The same calculations hold for the \emph{b}-quark.

With these results we find the probability that the top quark evolves
from scale $\tilde{q}_{i}^{2}$ to $\tilde{q}_{max}^{2}$ without
emitting any radiation is\begin{equation}
S\left(\tilde{q}_{i}^{2},\tilde{q}_{max}^{2}\right)=\frac{\Delta\left(\tilde{q}_{0}^{2},\tilde{q}_{max}^{2}\right)}{\Delta\left(\tilde{q}_{0}^{2},\tilde{q}_{i}^{2}\right)}.\label{eq:2.5.7}\end{equation}
Therefore the probability that there is some resolvable radiation
emitted as the top quark evolves from scales $\tilde{q}_{i}^{2}$
to $\tilde{q}_{max}^{2}$ is $1-S\left(\tilde{q}_{i}^{2},\tilde{q}_{max}^{2}\right)$,
which may be written as\begin{eqnarray}
-\int_{\tilde{q}_{i}^{2}}^{\tilde{q}_{max}^{2}}\mathrm{d}\tilde{q}^{2}\textrm{ }\frac{\mathrm{d}S\left(\tilde{q}_{i}^{2},\tilde{q}^{2}\right)}{\mathrm{d}\tilde{q}^{2}} & = & \int_{\tilde{q}_{i}^{2}}^{\tilde{q}_{max}^{2}}\mathrm{d}\tilde{q}^{2}\mathrm{d}z\textrm{ }\frac{\alpha_{\mathrm{S}}}{2\pi\tilde{q}^{2}}P_{qq}\left(z,\tilde{q}^{2}\right)S\left(\tilde{q}_{i}^{2},\tilde{q}^{2}\right)\Theta\left(\mathcal{R}\right).\label{eq:2.5.8}\end{eqnarray}
The integrand of (\ref{eq:2.5.8}) is the probability density that
the next resolvable branching after $\tilde{q}_{i}^{2}$ occurs at
scale $\tilde{q}^{2}$. 

The probability distribution (\ref{eq:2.5.8}) is sampled using the
veto algorithm as described in \cite{Gieseke:2003hm}. Finally we
note that, the gluons produced by the top quark and bottom quark will
in turn produce their own parton showers; these showers are standard
time-like gluon showers, not subject to any complications, \emph{e.g.}
initial-state evolution or parton masses, as is the case for the top
quark shower, the details of such gluon showering are described in
\cite{Gieseke:2003rz}.

\subsection{Phase-space limits - boundary conditions}

We will now discuss the integration limits for the Sudakov form factor
starting with $z$. The boundaries on the $z$ variable depend on
the $\tilde{q}^{2}$. For a given $\tilde{q}^{2}$ the $z$ boundaries
are determined by the requirement that $\mathbf{p}_{\perp}^{2}>0$.
In the case of the top quark the requirement that the intermediate
top quark has a mass greater than $\left(m_{b}+m_{W}\right)^{2}$
provides a further constraint on $z$. The implementation of these
complicated $z$ bounds is straightforward using the veto algorithm
\cite{Gieseke:2003hm}.

The lower bound on $\tilde{q}^{2}$ for both top and bottom quarks
is simply $\tilde{q}^{2}=m_{q}^{2}$ due to the fact that the top
quark is initially on-shell and the final \emph{b}-quark must also
be on-shell. The phase-space boundaries in the $\tilde{q}^{2}$ variable
were calculated in \cite{Gieseke:2003rz}. It was shown in \cite{Gieseke:2003rz},
that for a given $\tilde{q}_{b,max}$ there is an upper bound $\tilde{q}_{t,max}$,
for emissions from the top quark,\begin{equation}
\begin{array}{rcl}
\left(\tilde{q}_{t,max}^{2}-m_{t}^{2}\right)\left(\tilde{q}_{b,max}^{2}-m_{b}^{2}\right) & = & \frac{1}{4}\left(m_{t}^{2}-m_{W}^{2}+m_{b}^{2}+\lambda\right)^{2},\end{array}\label{eq:2.6.1}\end{equation}
 that partitions the phase-space for the decay $t\rightarrow bW+g$
into three regions which do not overlap: one region accessible to
gluon emissions from the bottom quark, one accessible to emissions
from the top quark, and a further inaccessible \emph{dead} \emph{region}. 

This partitioning is vital to ensure that each phase space point has
a unique matrix element approximation assigned to it and to avoid
double counting of phase-space points. As such the three regions do
not overlap, this is particularly significant for the soft gluon region
of phase space, which is most heavily populated. The price to be paid
for this partitioning is the presence of the dead region, however
this region is solely comprised of high $\mathbf{p}_{\perp}$ gluons,
so it is sparsely populated, and in any case the parton-shower approximation
is known to model such emissions badly. 

\begin{figure}[t]
\begin{center}\includegraphics[%
  clip,
  width=0.35\textwidth,
  keepaspectratio,
  angle=90]{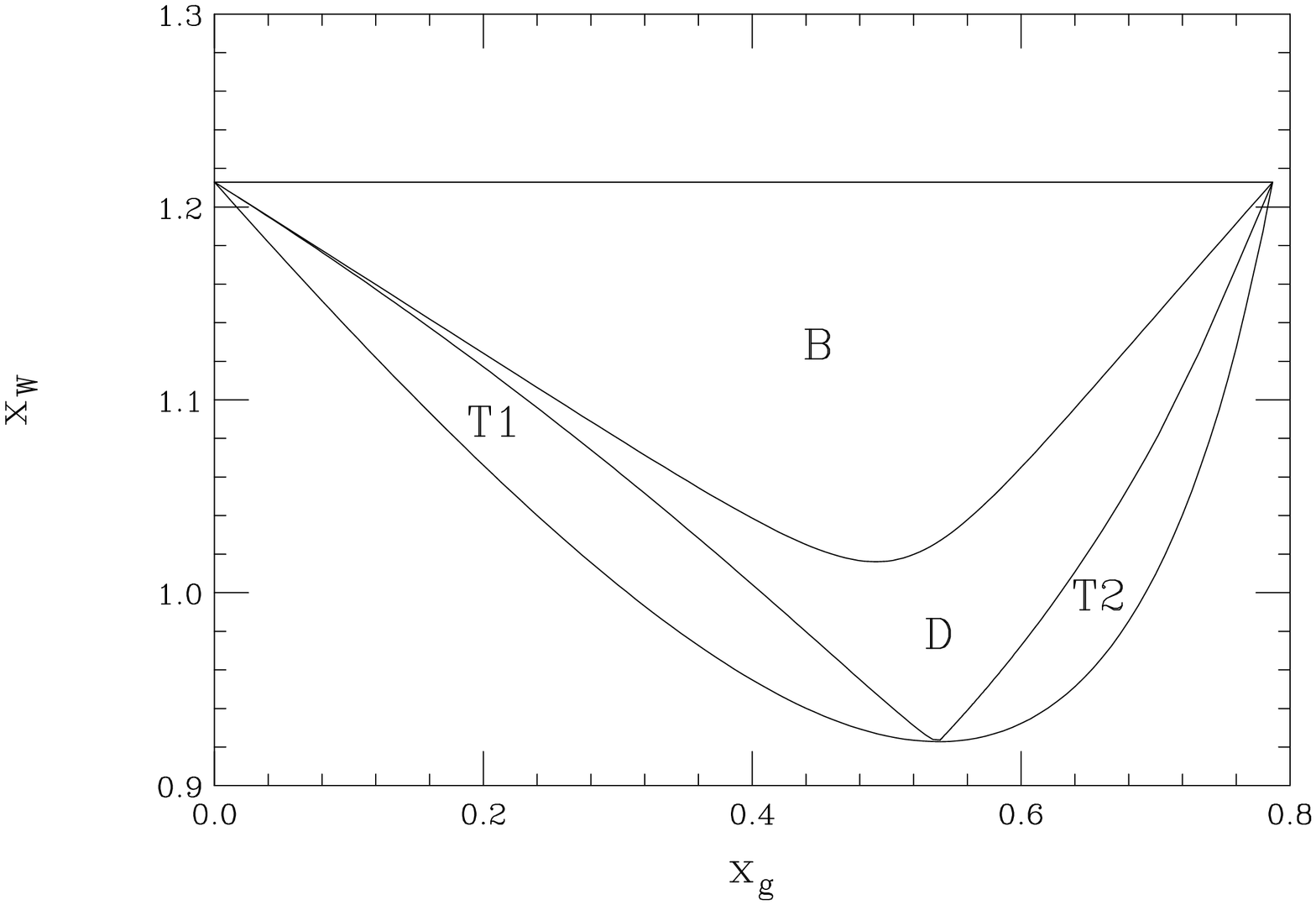}\hfill{}\includegraphics[%
  clip,
  width=0.35\textwidth,
  keepaspectratio,
  angle=90]{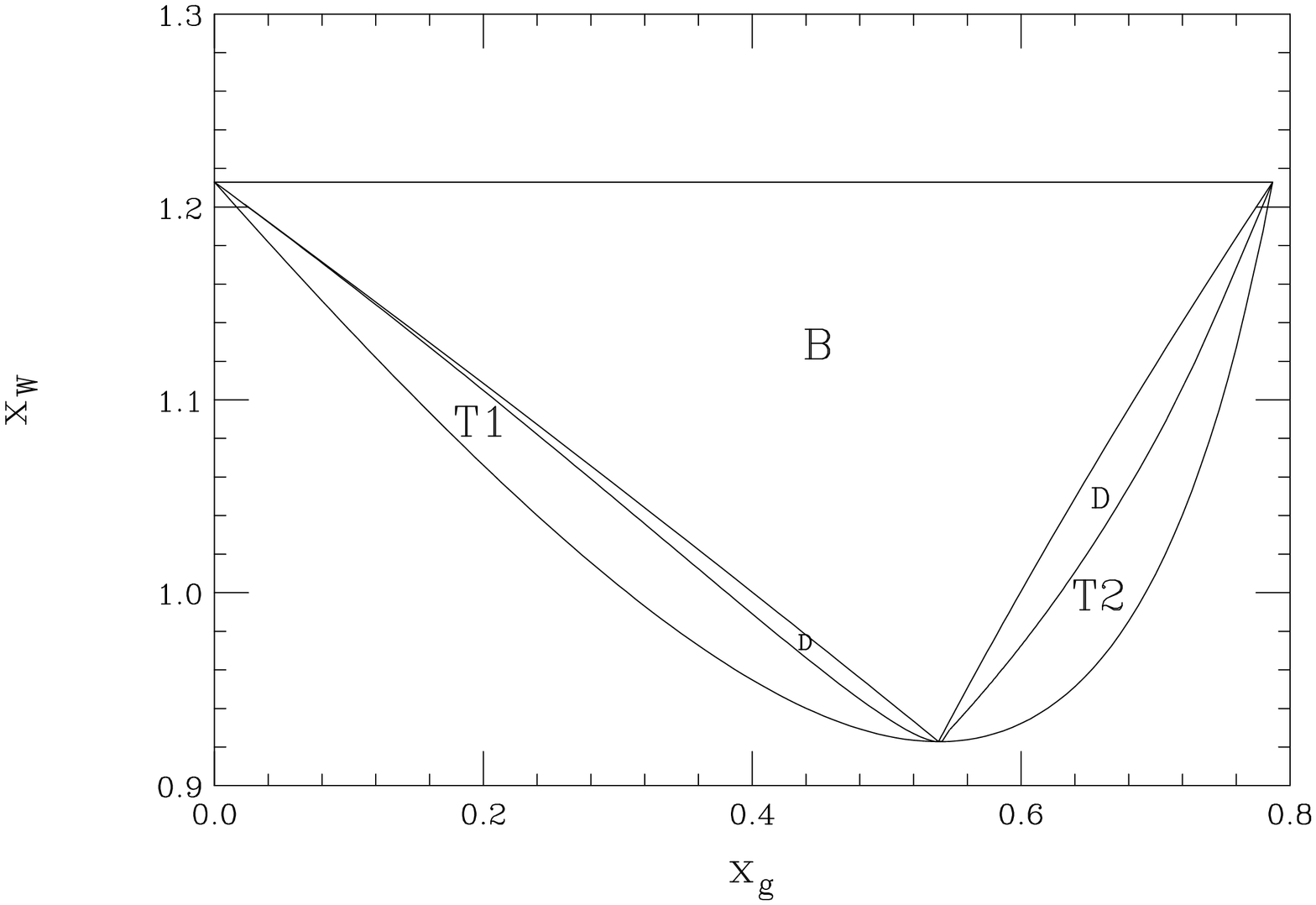}\end{center}

\caption{\label{fig:phase_space}In this figure we show the phase space boundaries
for the \emph{symmetric} (left) and \emph{maximal} (right) choices
of phase space partitioning, in the $x_{g},x_{W}$ plane, where $x_{g}$
and $x_{W}$ are equal to two times the energy fraction of the gluon
and \emph{W} boson in the top quark rest frame. The regions T1 and
T2 are populated by gluon emissions from the top quark while the region
labelled B is populated by emissions from the \emph{b}-quark. The
region labelled D is the \emph{dead region}. For the \emph{symmetric}
choice the phase space volume is divided more or less evenly between
that accessible to emissions from the top and bottom quarks, while
the \emph{maximal} choice maximises the volume to be populated by
emissions from the \emph{b}-quark\emph{.} }
\end{figure}

Given the relation (\ref{eq:2.6.1}), between $\tilde{q}_{t,max}^{2}$
and $\tilde{q}_{b,max}^{2}$, we define two choices of phase-space
partition, the so-called \emph{symmetric} choice,\begin{equation}
\begin{array}{lcl}
\tilde{q}_{b,max}^{2} & = & m_{b}^{2}+\frac{1}{2}\left(m_{t}^{2}-m_{W}^{2}+m_{b}^{2}+\lambda\right)\end{array}\label{eq:2.6.2}\end{equation}
and \emph{maximal} choice\begin{equation}
\begin{array}{lcl}
\tilde{q}_{b,max}^{2} & = & 4\left(\left(m_{t}-m_{W}\right)^{2}-m_{b}^{2}\right)\end{array}.\label{eq:2.6.3}\end{equation}
 In the case of the \emph{symmetric} choice, the phase-space volume
is divided more or less evenly between that accessible to emissions
from the top and bottom quarks, while the \emph{maximal} choice maximises
the volume to be populated by emissions from the \emph{b}-quark\emph{.}
The two choices of phase-space partitioning can be seen in figure
\ref{fig:phase_space}. Typically we favour the \emph{symmetric} choice,
as the \emph{maximal} choice involves generating high transverse momentum
emissions from the \emph{b}-quark. 

\subsection{Kinematics reconstruction\label{sub:What_the_code_actually_does.}}

Given the boundary conditions and the emission probability distribution
(\ref{eq:2.5.8}) we have all we need to generate parton showers in
terms of $\tilde{q}_{i}^{2}$, $z_{i}$ and $\phi_{i}$. The Sudakov
variable $\alpha_{i}$ may be calculated from (\ref{eq:2.1.3}) as
each $z_{i}$ is generated, the transverse momentum is also calculated
at this point. The calculation of the $\beta_{i}$ variables is more
complicated. 

For time-like, final-state, showers we start at the end of the shower\emph{,}
since the particles are on-shell there we can simply determine the
value of $\beta_{i}$ by computing $q_{i}^{2}$ from (\ref{eq:2.1.1})
and setting $q_{i}^{2}=m_{q}^{2}$, where $m_{q}^{2}$ is the on-shell
mass squared. Once this is done the \emph{parent} particle's virtuality,
and hence its $\beta_{i}$ value, follows directly from momentum conservation.
Applying momentum conservation to each vertex in such showers, one
can fully reconstruct all of the momenta up to and including that
of the \emph{shower} \emph{progenitor} \emph{particle}. 

Recall that the gluons radiated by the top quark will produce their
own time-like showers. The first step in reconstructing the momenta
of the top quark as it evolves toward its decay, is to reconstruct
these showers, \emph{i.e.} we first reconstruct the momenta of the
gluons that were radiated by the top quark by applying the procedure
outlined above. The momentum of the top quark after each gluon emission
can then be calculated by momentum conservation, as the initial top
quark momentum is known $\left(m_{t},\mathbf{0},0\right)$. In this
way we completely determine the momenta of the top quark and all of
the radiation emitted prior to the $t\rightarrow bW$ decay. Should
the \emph{b}-quark also emit radiation, the resulting jet(s) will
be reconstructed in the standard way for time-like showers \cite{Gieseke:2003hm}. 

Since the initial \emph{b} and \emph{W} boson momenta were generated
according to the tree level $t\rightarrow bW$ decay they add up to
$\left(m_{t},\mathbf{0},0\right)$, rather than the actual momentum
of the top quark prior to its decay. Furthermore, if the \emph{b}-quark
radiates, forming a \emph{b-}jet%
\footnote{In the present discussion we will assume the \emph{b}-quark has given
rise to a jet since this is the general case, the extension to the
case where it does not radiate is trivial. %
} the jet reconstruction described above leads to the initial \emph{b}-quark
momentum having a virtuality greater than its mass. 

\begin{figure}[t]
\begin{center}\includegraphics[%
  clip,
  width=0.50\textwidth,
  keepaspectratio]{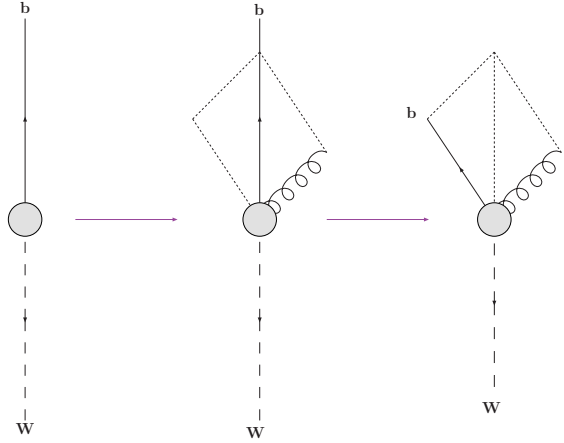}\end{center}

\caption{\label{fig:reshuffling}In this figure we sketch the `momentum reshuffling'
procedure. Initially the top quark decay to a \emph{b}-quark and a
\emph{W} boson is simulated, this momentum configuration is shown
first on the far left. Afterwards some additional radiation is produced
from the parton shower, represented by the spiral in the central configuration
above, clearly this configuration does not conserve momentum. The
initial \emph{b-}jet momenta are rescaled and boosted to give the
configuration shown on the right hand side: the three-momenta of the
\emph{W} boson and \emph{b}-jet are scaled down and the momentum of
the additional radiation, transverse to the \emph{W}-boson direction,
is absorbed by the \emph{b-}jet .}
\end{figure}

In order to ensure global momentum conservation we perform a `momentum
reshuffling' which smoothly preserves the internal properties of each
jet. Let us denote the total momentum of the jets radiated by the
top quark $g_{ISR}$. The first step in the momentum reshuffling involves
rescaling the three-momenta of the \emph{W} boson to account for the
loss of energy due to gluon emissions from the top quark. The second
step involves absorbing the component of $g_{ISR}$ transverse to
the \emph{W}, in the \emph{b}-jet. Finally the momentum of the \emph{b}-jet
in the direction of the original \emph{b}-quark is rescaled. This
process is sketched in figure \ref{fig:reshuffling} and is described
by the following mapping:{\small \begin{equation}
\begin{array}{lcl}
q_{W} & = & \left(\sqrt{m_{W}^{2}+\mathbf{p}^{2}},-\mathbf{p}\right)\rightarrow\left(\sqrt{m_{W}^{2}+k_{2}^{2}\mathbf{p}^{2}},-k_{2}\mathbf{p}\right),\\
q_{b_{JET}} & = & \left(\sqrt{m_{b_{JET}}^{2}+\mathbf{p}_{b_{JET}}^{2}},\mathbf{p}_{b_{JET}}\right)\rightarrow\left(\sqrt{m_{b_{JET}}^{2}+k_{1}^{2}\mathbf{p}^{2}+\mathbf{p}_{ISR\perp}^{2}},k_{1}\mathbf{p}-\mathbf{p}_{ISR\perp}\right),\end{array}\label{eq:2.7.1}\end{equation}
}where $\mathbf{p}$ is the initial \emph{W} boson momentum from the
$t\rightarrow bW$ process, $k_{1}$ and $k_{2}$ are constant rescalings
of the \emph{b}-jet and \emph{W} boson momenta. The three vectors
$\mathbf{p}_{ISR\parallel}$ and $\mathbf{p}_{ISR\perp}$ are the
components of $g_{ISR}$ parallel and perpendicular to $\mathbf{p}$.
Applying energy-momentum conservation to the momenta above allows
one to determine the value of the rescalings $k_{1}$ and $k_{2}$.
The top quark and the radiation from it are untouched by the momentum
reshuffling.

In practice, rather than applying the rescalings and momentum subtractions
as indicated in (\ref{eq:2.7.1}), once $k_{1}$ and $k_{2}$ are
determined we actually calculate the Lorentz \emph{}boosts \emph{}which
perform a mapping that is exactly equivalent to that in (\ref{eq:2.7.1}).
By calculating the mapping in terms of these Lorentz boosts, we can
conserve momentum and preserve the internal structure of the jets
by applying the boosts to each particle in the \emph{b-}quark jet. 

If the top quark does not radiate any gluons before decaying, the
three-momenta of the \emph{b}-jet and \emph{W} boson system are rescaled
such that its invariant mass is equal to the top mass. In this special
case the problem of momentum reshuffling is identical to that of $e^{+}e^{-}\rightarrow q\bar{q}$
events, hence for these cases we use exactly the same momentum reshuffling
as described in \cite{Gieseke:2003hm}. 

This sequence of boosts and rescalings is designed such that, at least
for the case of one gluon emission, the 3-body decay kinematics of
\cite{Gieseke:2003rz} are reproduced, \emph{i.e.} should the gluon
be emitted by the top quark, or the \emph{b-}quark, \emph{}its transverse
momentum is absorbed by the \emph{b}-quark, in the top quark rest
frame (as depicted in figure \ref{fig:reshuffling}).

\section{Matrix element corrections\label{sec:Matrix-Element-Corrections}}

As stated in section \ref{sec:Showering}, the effects of unresolvable
gluon emissions have been included to all orders through the Sudakov
form factor. The master formula and shower algorithms generate further
resolvable emissions by approximating the full $t\rightarrow bW+\left(n\right)g$
matrix element by a product of quasi-collinear splitting functions
multiplying the tree level amplitude. Ideally, we wish to include
the higher-order effects in the master equation as accurately as possible.

\subsection{Soft matrix element corrections\label{sub:Soft-Matrix-Element-Corrections}}

In the parton shower approximation the probability density that the
$i$th resolvable gluon is emitted into $\left[\tilde{q}^{2},\tilde{q}^{2}+\mathrm{d}\tilde{q}^{2}\right],$
$\left[z,z+\mathrm{d}z\right]$ is \begin{equation}
dP\left(z,\tilde{q}^{2}\right)=\frac{\mathrm{d}\tilde{q}^{2}}{\tilde{q}^{2}}\mathrm{d}z\textrm{ }\frac{\alpha_{\mathrm{S}}\left(\mathbf{p}_{T}\right)}{2\pi}P_{qq}\left(z,\tilde{q}^{2}\right)\Theta\left(\mathcal{R}\right).\label{eq:3.1.1}\end{equation}
This approximation works well for the case that the emission lies
within the domain of the quasi-collinear limit. On the other hand
the exact matrix element calculation gives us that the probability
of a resolved emission is (at least in the leading log approximation)
\begin{equation}
\int_{\mathcal{R}}\mathrm{d}P^{\mathrm{m}.\mathrm{e}.}=\int\mathrm{d}\tilde{q}^{2}\mathrm{d}z\textrm{ }\frac{1}{\Gamma_{0}}\frac{\mathrm{d}^{2}\Gamma}{\mathrm{d}z\mathrm{d}\tilde{q}^{2}}\Theta\left(\mathcal{R}\right),\label{eq:3.1.2}\end{equation}
where $\Gamma$ is the width of the process $t\rightarrow bWg$. The
differential width for $t\rightarrow bWg$ is given in appendix \ref{sub:Matrix-element.}.
Once again the `probability' of an unresolved emission can therefore
be written $-\int_{\mathcal{R}}\mathrm{d}P^{\mathrm{m}.\mathrm{e}}$,
proceeding in the same way as our earlier derivations (\ref{eq:2.5.8}),
we then have the probability density that the $i$th resolvable gluon
is emitted into $\left[\tilde{q}^{2},\tilde{q}^{2}+\mathrm{d}\tilde{q}^{2}\right]$,
$\left[z,z+\mathrm{d}z\right]$ is given by the integrand of \begin{equation}
\int_{\tilde{q}_{i-1}^{2}}^{\tilde{q}_{max}^{2}}\mathrm{d}\tilde{q}_{i}^{2}\mathrm{d}z\textrm{ }\frac{1}{\Gamma_{0}}\frac{\mathrm{d}^{2}\Gamma}{\mathrm{d}z\mathrm{d}\tilde{q}_{i}^{2}}\textrm{ }\exp\left(-\int_{\tilde{q}_{i-1}^{2}}^{\tilde{q}_{i}^{2}}\mathrm{d}\tilde{q}^{2}\mathrm{d}z\textrm{ }\frac{1}{\Gamma_{0}}\frac{\mathrm{d}^{2}\Gamma}{\mathrm{d}z\mathrm{d}\tilde{q}^{2}}\right).\label{eq:3.1.3}\end{equation}
We may generate the distribution in (\ref{eq:3.1.3}) by simply augmenting
the veto algorithm that is used to produce (\ref{eq:2.5.8}) with
a single additional rejection weight, simply vetoing emissions if
a random number $\mathcal{R}_{S}$ is such that\begin{equation}
\mathcal{R}_{S}\ge\left.\frac{\mathrm{d}P}{\mathrm{d}P}^{\mathrm{m}.\mathrm{e}.}\right|_{z,\tilde{q}^{2}}.\label{eq:3.1.4}\end{equation}
For this to work we require that the parton shower emission probability
$\mathrm{d}P$ always overestimates that of the exact matrix element
$\mathrm{d}P^{\mathrm{m}.\mathrm{e}.}$, if necessary this can be
achieved by simply enhancing the emission probability of the parton
shower with a constant factor. 

In practice we do not apply this correction to all emissions as most
should be well approximated by the parton shower. The emissions which
are not expected to be modelled well are neither soft nor collinear,
they have large transverse momentum $\left(\mathbf{p}_{\perp}\right)$.
In practice we make the ansatz that all emissions which do not have
the largest $\mathbf{p}_{\perp}$ of any generated thus far, are considered
to be infinitely soft and we only correct the set which is comprised
of those emissions which have the largest $\mathbf{p}_{\perp}$ \emph{so}
\emph{far}, to the full matrix element distribution. 

One might be concerned that in this case it is really only proper
to apply this correction to the final, largest $\mathbf{p}_{\perp}$emission,
however, in the context of our coherent parton branching formalism
(angular ordering) the earlier wide angle emission is considered too
soft to resolve the subsequent, larger $\mathbf{p}_{\perp}$ splitting,
and is therefore effectively distributed assuming that the latter
emission did not occur. Under these assumptions the correct procedure
should be to correct only those emissions which are the hardest so
far, from distribution (\ref{eq:3.1.1}) to distribution (\ref{eq:3.1.3})
by applying veto in (\ref{eq:3.1.4}) \cite{Seymour:1994df}.

\subsection{Hard matrix element corrections\label{sub:Hard-Matrix-Corrections}}

In addition to correcting the distribution of radiation inside the
T1, T2 and B regions populated by the parton shower, we also wish
to correct the distribution of radiation outside, in the high $\mathbf{p}_{\perp}$
dead region. We wish to distribute the radiation in the dead region
according to the full $t\rightarrow bWg$ matrix element \emph{i.e.}
according to\begin{equation}
\frac{1}{\Gamma_{0}}\frac{\mathrm{d}^{2}\Gamma}{\mathrm{d}x_{g}\mathrm{d}x_{W}}=\frac{\alpha_{S}C_{F}}{\pi}\left(\frac{f\left(b,w,\tilde{q}^{2},z\right)}{\lambda\left(1+w-b-x_{W}\right)x_{g}^{2}}\right),\label{eq:3.2.1}\end{equation}
where $b=m_{b}^{2}/m_{t}^{2}$, $w=m_{W}^{2}/m_{t}^{2}$ and $\lambda$
is given by (\ref{eq:2.1.2.b}). The full expression for the width
is lengthy and so we give it in Appendix \ref{sub:Matrix-element.}. 

The first step in the algorithm is to generate a point inside the
dead region. This is non-trivial as the phase-space boundaries shown
in figure \ref{fig:phase_space} are rather complicated functions
of the mass of the \emph{b}-quark, \emph{W}-boson, $z$ and $\tilde{q}^{2}$.
It turns out that the phase-space may be parametrized as functions
$x_{W}\left(x_{g},\tilde{q}^{2}\right)$, that is, the Dalitz variable
$x_{W}=2q_{W}.p_{t}/m_{t}^{2}$ may be written as a function of $x_{g}=2q_{g}.p_{t}/m_{t}^{2}$
and the evolution variable $\tilde{q}^{2}$(to do this one eliminates
$z$ \emph{}in favour of $x_{g}$). These functions were calculated
using the conventions in \cite{Gieseke:2003rz} and are noted in appendix
\ref{sub:Phase-space}.

The phase-space point is selected by importance sampling the differential
distribution (\ref{eq:2.3.1}) assuming a $x_{g}^{-\alpha}\left(1+w-x_{W}\right)^{-1}$
behaviour, where $\alpha$ is a parameter which may be tuned to improve
the sampling efficiency. First we rewrite the integral over the dead
region $\left(\mathrm{D}\right)$ as,\begin{equation}
\begin{array}{rcl}
\frac{1}{\Gamma_{0}}\int_{\mathrm{D}}\mathrm{d}x_{g}\mathrm{d}x_{W}\textrm{ }\frac{\mathrm{d}^{2}\Gamma}{\mathrm{d}x_{g}\mathrm{d}x_{W}} & = & \frac{1}{\Gamma_{0}}\int_{\mathrm{D}}\mathrm{d}y_{g}\mathrm{d}y_{W}\textrm{ }\mathcal{W}\left(x_{g},x_{W}\right),\\
\mathcal{W}\left(x_{g},x_{W}\right) & = & \textrm{ }x_{g}^{\alpha}\left(1+w-x_{W}\right)\frac{\mathrm{d}^{2}\Gamma}{\mathrm{d}x_{g}\mathrm{d}x_{W}},\\
y_{g} & = & \frac{1}{1-\alpha}x_{g}^{1-\alpha},\\
y_{W} & = & \ln\left(1+w-x_{W}\right),\end{array}\label{eq:3.2.2}\end{equation}
and generate $y_{g},y_{W}$ points assuming they are uniformly distributed
between their maximum and minimum values in the dead region. Then
we keep events with probability\begin{equation}
\begin{array}{lcl}
\mathcal{P} & = & \frac{\mathcal{W}\left(x_{g},x_{W}\right)\mathcal{V}\left(y_{g}\right)}{\Gamma_{0}},\end{array}\label{eq:3.2.3}\end{equation}
$\left(\mathcal{P}\le1\right)$ where \begin{equation}
\mathcal{V}\left(y_{g}\right)=\left(y_{g,max}-y_{g,min}\right)\left.\left(y_{W,max}-y_{W,min}\right)\right|_{y_{g}},\label{eq:3.2.4}\end{equation}
 is the Monte Carlo estimate of the volume of the dead region in $y_{g},y_{W}$
plane: $y_{g,max}$ and $y_{g,min}$ are the maximum and minimum possible
values for $y_{g}$ in the dead region, while $y_{W,max}$ and $y_{W,min}$
are the maximum and minimum possible values of $y_{W}$ in the dead
region, for the generated value of $y_{g}$. 

Once a pair of values $y_{g},\textrm{ }y_{W}$ has been successfully
generated, they may be mapped back to the corresponding $x_{g},\textrm{ }x_{W}$
phase-space point and the momenta of the top quark, bottom quark and
W boson can be constructed from there.

\section{Results\label{sec:Results}}

\begin{figure}[t]
\begin{center}\includegraphics[%
  clip,
  width=0.35\textwidth,
  keepaspectratio,
  angle=90]{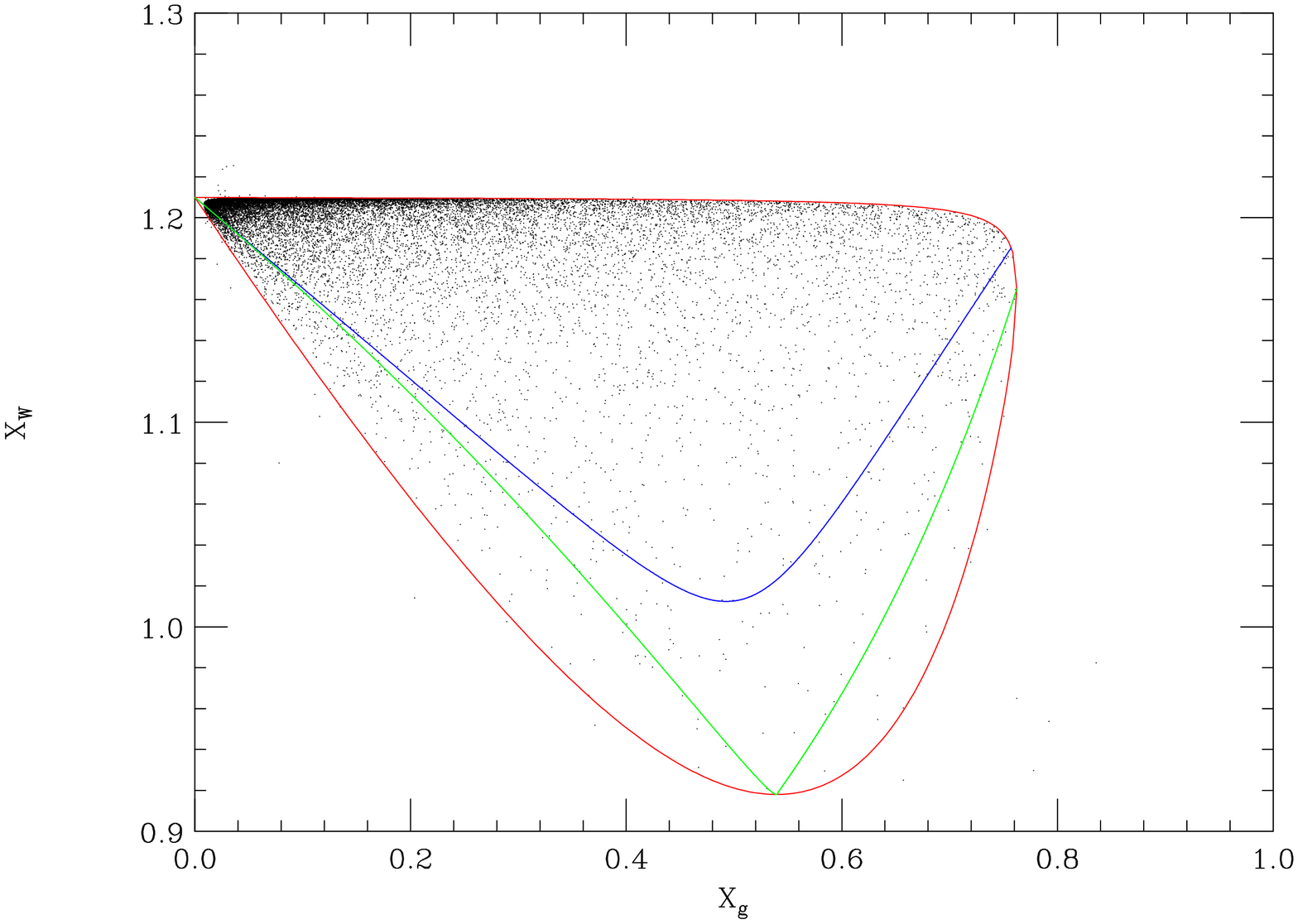}\hfill{}\includegraphics[%
  clip,
  width=0.35\textwidth,
  keepaspectratio,
  angle=90]{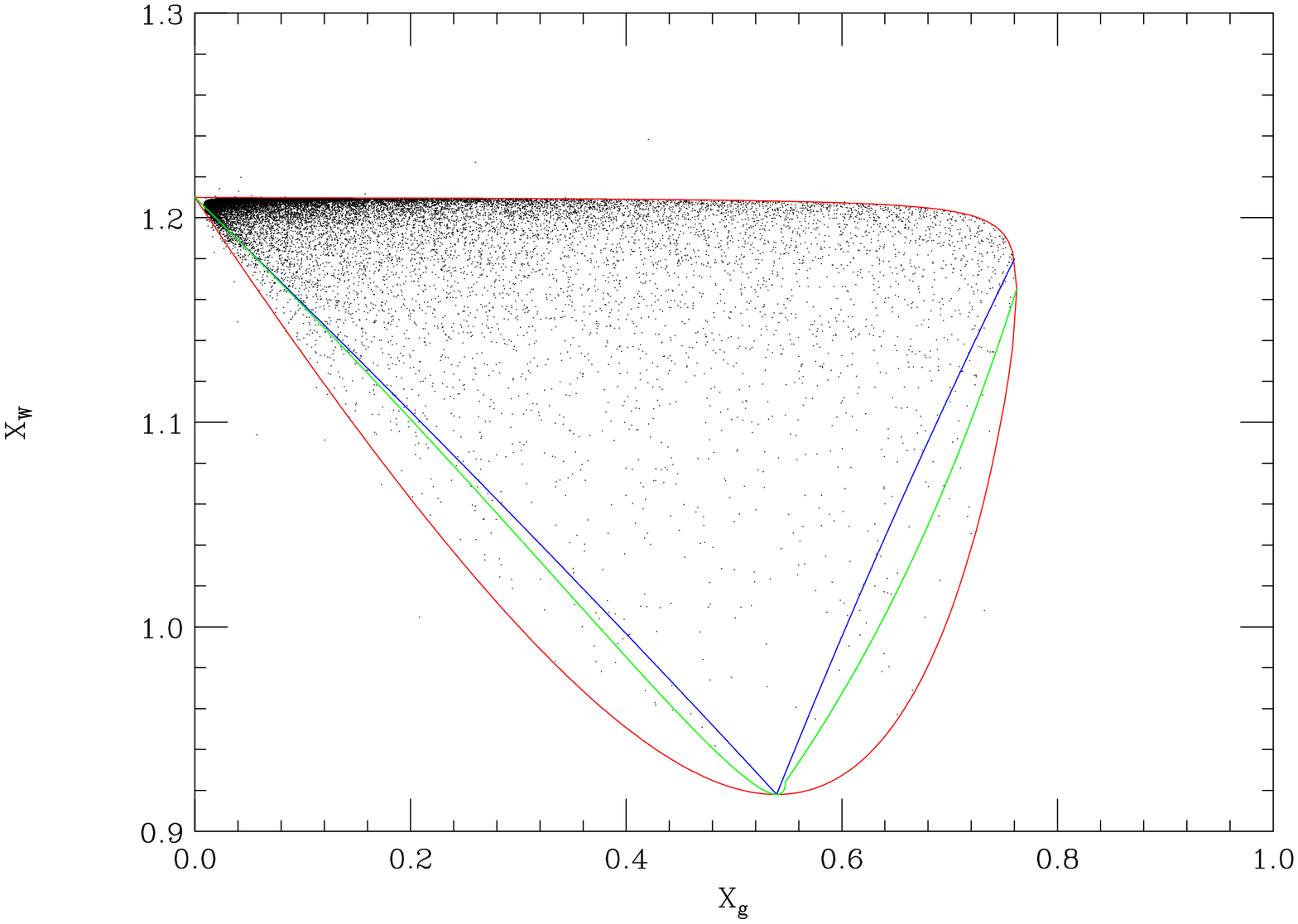}\end{center}

\caption{\label{fig:dalitz_population} Dalitz plot for gluonic radiation
in top decay. In both plots the soft and hard matrix element corrections
have been applied, but only one emission has been allowed. a) shows
the radiation for the symmetric choice of \cite{Gieseke:2003rz} for
emission from the top and bottom while b) shows the radiation with
the scales chosen to give the maximum amount of radiation from the
bottom quark. The blue (innermost) line gives the limit for radiation
from the bottom, the green (middle) line from the top and the red
(outer) line the boundary of the phase space region.}
\end{figure}

In figure \ref{fig:dalitz_population} we show the Dalitz distributions
for the process $t\rightarrow bWg$ as given by the parton shower
algorithm, including matrix element corrections. Only decays where
\emph{one} gluon is emitted are considered, since to do otherwise
would constitute a different (higher dimensional) phase-space volume
to that shown. 

The Dalitz plots show significant clustering of events in the soft
$x_{g}\rightarrow$0 region as expected. One can also see a marked
clustering of events along the line $x_{W}\approx\textrm{1.2}$ where
the energy fraction of the \emph{W} boson is \emph{maximal} \emph{i.e.}
where the gluon is emitted collinear with the \emph{b-}quark. Again,
this is to be expected since we know that collinear emissions from
the \emph{b-}quark are enhanced by logarithms of $m_{b}^{2}/m_{t}^{2}$.
We do not see a similar wedge-like clustering of events at low transverse
momentum with respect to the top quark. Such events would lie along
the lower section of the red boundary between $x_{g}\approx\textrm{0}$
and $x_{g}\approx\textrm{0.75}$. Clustering along this line would
only come from collinear enhancements of the form $\log\frac{Q^{2}}{m_{t}^{2}}$
where $Q^{2}$ is some hard scale, but in this case $Q^{2}=m_{t}^{2}$
so there is no such enhancement, only soft enhancements $\log\frac{m_{t}^{2}}{m_{g}^{2}}$
giving rise to the observed clustering as $x_{g}\rightarrow\textrm{0}$. 

In addition to considering the isolated decay process we have also
considered the process $e^{+}e^{-}\rightarrow t\bar{t}$ near threshold:
$\sqrt{s}=\textrm{360 GeV}$. By working close to the $t\bar{t}$
threshold we inhibit the effects of radiation from the production
phase of the $t\bar{t}$ pair, thus highlighting the effects of radiation
in the $t\bar{t}$ decays. In analysing these events we have worked
at the parton level and only considered leptonic $W$ decays. We have
clustered all final-state quarks and gluons into three jets using
the $k_{\perp}$ clustering algorithm \cite{Butterworth:2002xg},
taking care to omit the $W$ decay products from the clustering. Events
for which the minimum jet separation is less than $\Delta R=0.7$
$\left(\Delta R^{2}=\Delta\eta^{2}+\Delta\phi^{2}\right)$, as well
as events containing a jet with transverse energy less than 10 GeV,
are excluded from the analysis. For these events we have plotted the
distributions of the jet separation $\Delta R$ and the logarithm
of the jet measure $y_{3}$, where \begin{equation}
y_{3}=\frac{2}{s}\textrm{ }\textrm{min}_{ij}\left(\min\left(E_{i}^{2},E_{j}^{2}\right)\left(1-\cos\theta_{ij}\right)\right)\label{eq:4.1}\end{equation}

\noindent is the value of the jet resolution parameter for which the
three jet event would be seen as a two jet event. This analysis is
the same as that performed in two earlier related works \cite{Corcella:1998rs,Orr:1996pe}. 

\begin{figure}[t]
\begin{center}\includegraphics[%
  clip,
  width=0.35\textwidth,
  keepaspectratio,
  angle=90]{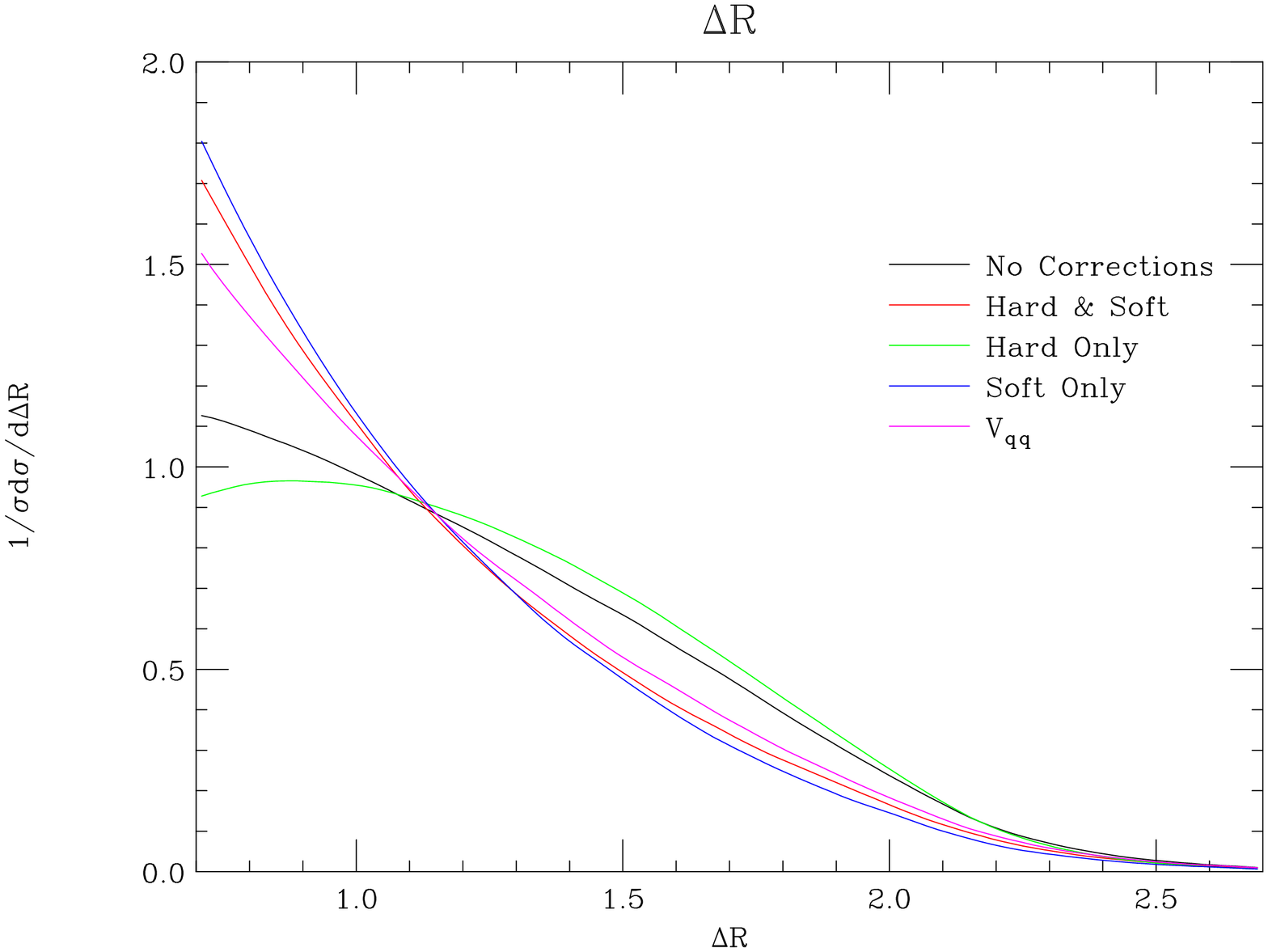}\hfill{}\includegraphics[%
  clip,
  width=0.35\textwidth,
  keepaspectratio,
  angle=90]{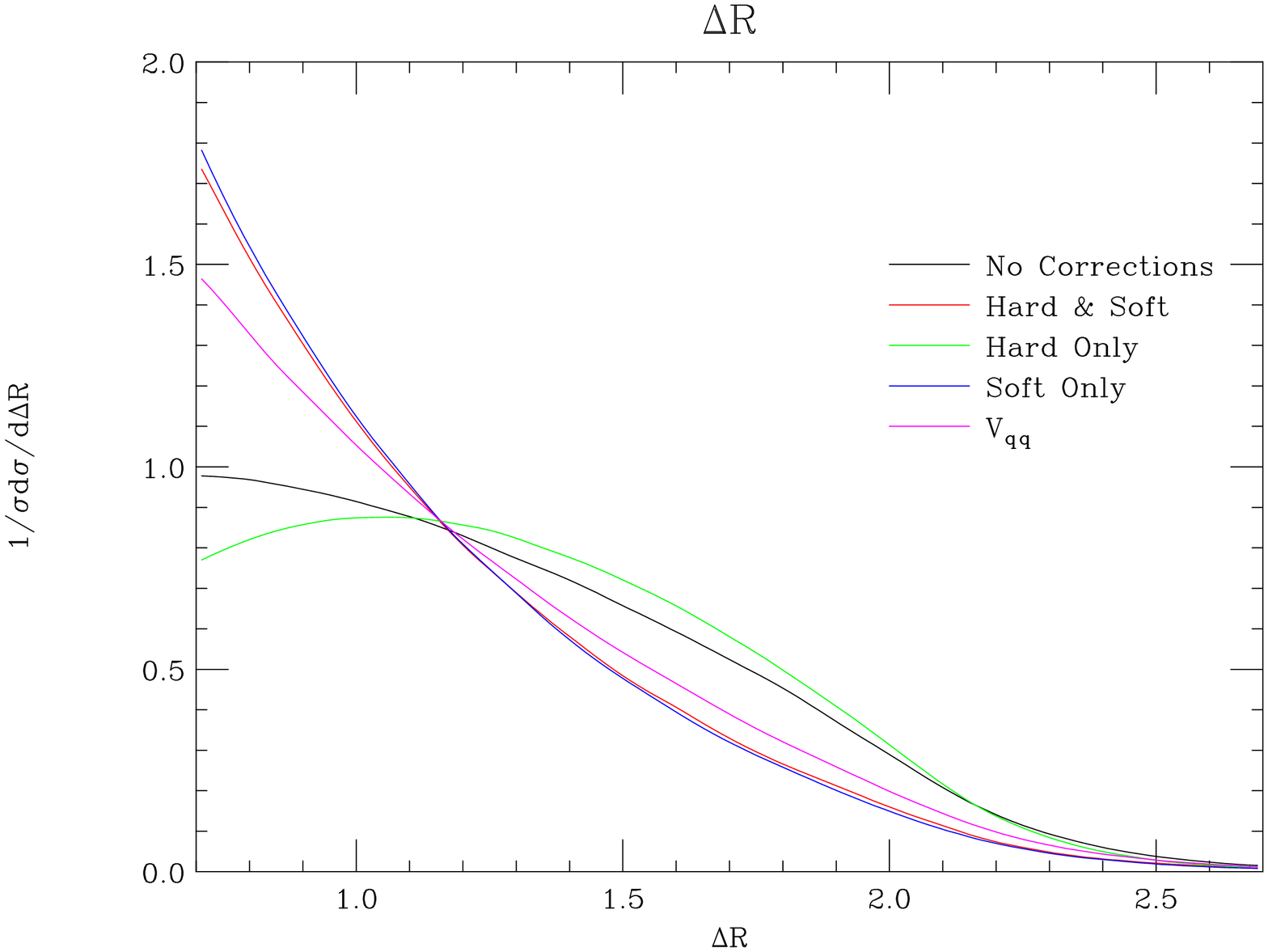}\end{center}

\caption{\label{fig:DeltaR}$\frac{1}{\sigma}\frac{\mathrm{d}\sigma}{\mathrm{d}\Delta R}$
where $\Delta R^{2}=\Delta\eta^{2}+\Delta\phi^{2}$ for three jet
$e^{+}e^{-}\rightarrow t\bar{t}$ events at $\sqrt{s}=360\mathrm{GeV}$with
and without matrix element corrections. On the left we show the distributions
obtained for the case that the phase-space volume populated by the
parton showers from the top and bottom quarks is almost the same size
(the \emph{symmetric} phase space partition (\ref{eq:2.6.2})), while
on the right we show the distributions obtained for the case that
the shower from the \emph{b-}quark \emph{}populates \emph{}most of
the phase space choice (the \emph{maximal} phase space partition (\ref{eq:2.6.3})).
In each plot the black line corresponds to the parton shower approximation,
the red line corresponds to the parton shower including hard and soft
matrix element corrections, while the green/blue lines respectively
correspond to including only the hard/soft part of the matrix element
corrections. The magenta line is obtained using the standalone parton
shower but with $P_{qq}$ replaced by the generalized quasi-collinear
splitting function $\mathcal{V}_{qq}$ (\ref{eq:2.4.6}).}
\end{figure}

In figure \ref{fig:DeltaR} we show differential distributions with
respect to the jet separation of the closest pair of jets in the event
$\left(\Delta R\right)$, for the cases of the \emph{symmetric} and
\emph{maximal} phase space partitions (\ref{eq:2.6.2}, \ref{eq:2.6.3}).
Figure \ref{fig:DeltaR} shows that the matrix element corrections
have significant consequences for the $\Delta R$ distributions. We
see that when the soft matrix element correction is applied, the distribution
is more peaked for small $\Delta R$ and tails off more quickly than
the distributions obtained without it. This is to be expected, it
is a \emph{softening} of the distribution, it is indicative of the
fact that the soft matrix element correction is vetoing a number of
high $p_{T}$ emissions, not well modelled by the standalone parton
shower: such hard emissions naturally give rise to more widely separated
jets (larger $\Delta R$). 

The degree to which the soft matrix element correction affects the
$\Delta R$ distribution suggests that the standalone parton shower
does not model the number of these high $p_{T}$ emissions well: similar
distributions were observed in version 5.9 of the older HERWIG \textsf{}program
and were since considered to be the result of a bug (showering in
the wrong reference frame). In our case the effect shown is understood
to be a genuine artefact of the covariant parton shower formalism
and may be traced back to the form of the quasi-collinear splitting
function for \emph{b-}quark emissions. In section \ref{sub:Matrix-element-approximations.}
we noted that, as the angle between the reference vector $\left(n\right)$
and the emitted gluon $\left(k\right)$ decreases, emissions will
be enhanced in the $n$ direction (particularly soft emissions). This
spurious enhancement results from choosing $n$ not equal to precisely
the momentum of the \emph{}colour partner of the \emph{b}-quark. This
also explains why the differences between the results with and without
matrix element corrections are more pronounced for the \emph{maximal}
phase space partition than the \emph{symmetric} partition. On the
contrary our results including the soft matrix element corrections
compare well with those obtained in the earlier \textsf{FORTRAN} HERWIG
program.

In section \ref{sub:Matrix-element-approximations.} we also proposed
a new generalized quasi-collinear splitting function $\mathcal{V}_{qq}$
(see also appendix \ref{sec:Generalization_of_Pqq}), which improves
on the usual quasi-collinear splitting function by reducing to the
eikonal dipole function in the soft limit. From a practical point
of view the introduction of this splitting function is akin to a soft
matrix element correction. The $\Delta R$ distributions support this
assertion; the standalone parton shower is greatly improved by the
use of this new splitting function. 

\begin{figure}[t]
\begin{center}\includegraphics[%
  clip,
  width=0.35\textwidth,
  keepaspectratio,
  angle=90]{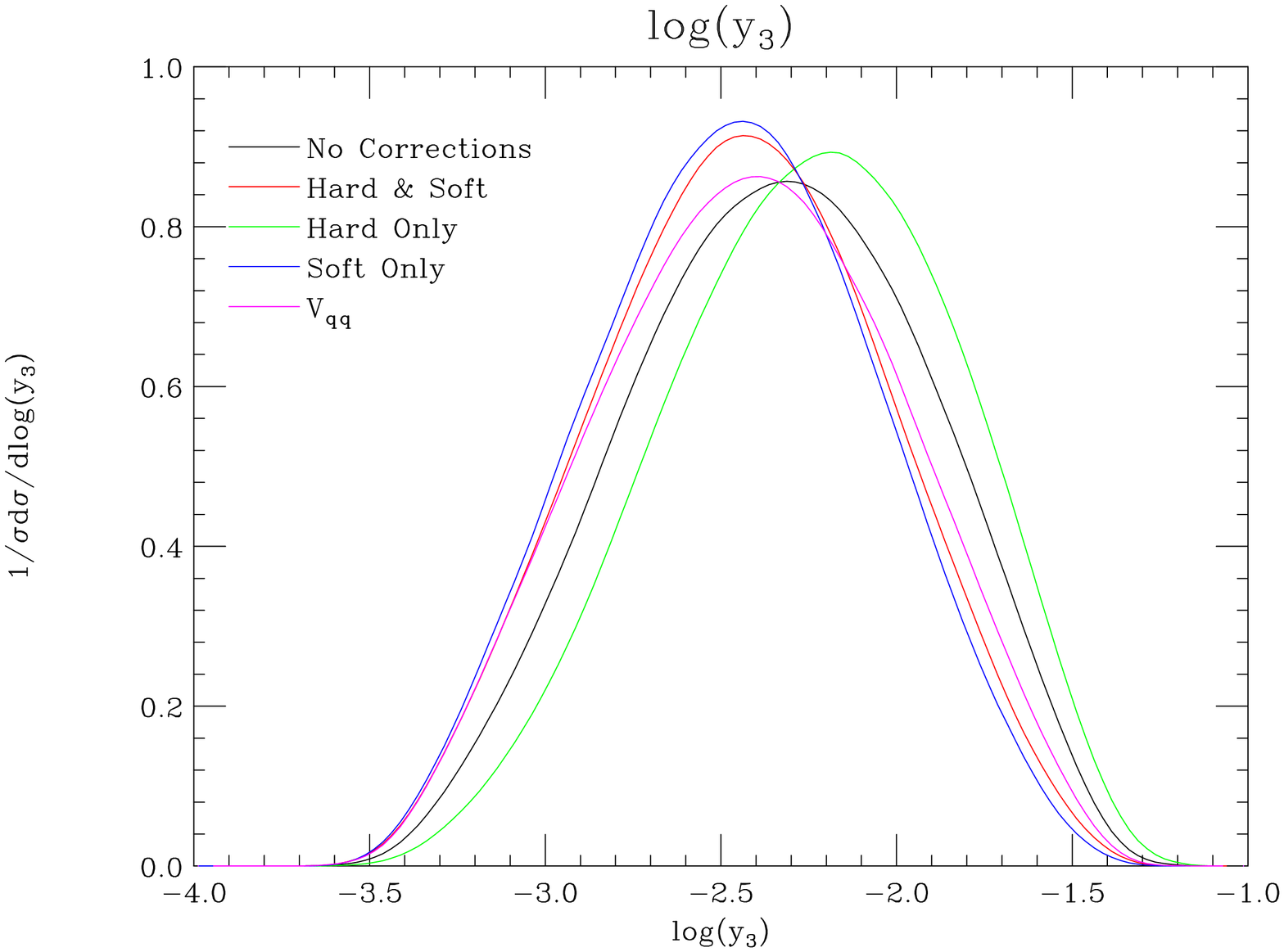}\hfill{}\includegraphics[%
  clip,
  width=0.35\textwidth,
  keepaspectratio,
  angle=90]{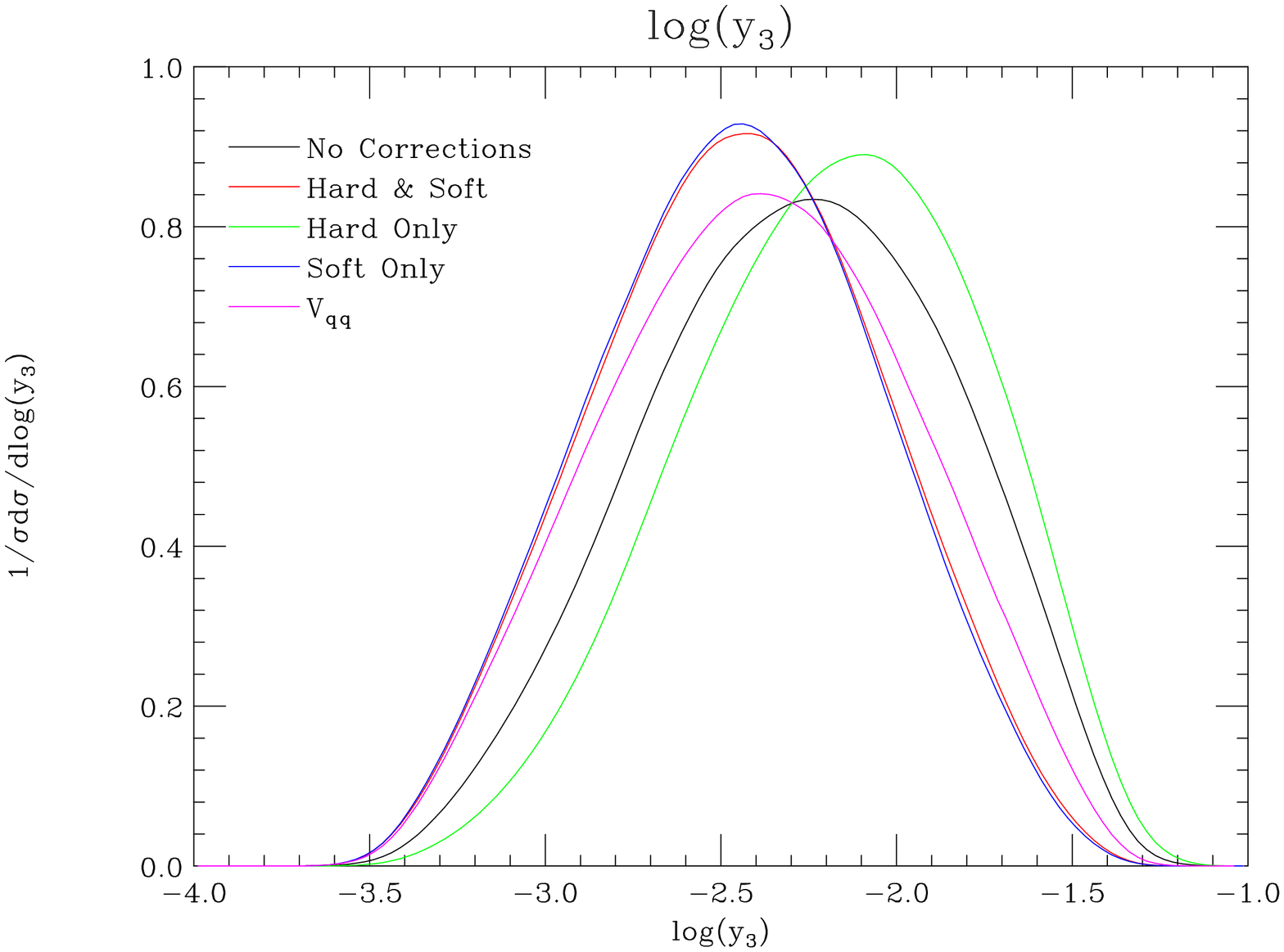}\end{center}

\caption{\label{fig:logy3}$\frac{1}{\sigma}\frac{\mathrm{d}\sigma}{\mathrm{d}\log\left(y_{3}\right)}$
where $y_{3}=\frac{2}{s}\textrm{ }\textrm{min}_{ij}\left(\min\left(E_{i}^{2},E_{j}^{2}\right)\left(1-\cos\theta_{ij}\right)\right)$
for three jet $e^{+}e^{-}\rightarrow t\bar{t}$ events at $\sqrt{s}=360\mathrm{GeV}$with
and without matrix element corrections. As in figure \ref{fig:DeltaR},
on the left we show the distributions obtained for the \emph{symmetric}
phase space partition (\ref{eq:2.6.2}), while on the right we show
the distributions obtained for the \emph{maximal} phase space partition
(\ref{eq:2.6.3}. The histograms are coloured in the same way as for
figure \ref{fig:DeltaR}.}
\end{figure}

As with figure \ref{fig:DeltaR}, in figure \ref{fig:logy3} we show
differential distributions with respect to the jet measure $y_{3}$,
for the cases of the \emph{symmetric} and \emph{maximal} phase space
partitions, for all possible combinations of matrix element corrections.
We also show the same distribution obtained using the generalized
quasi-collinear splitting function in the basic parton shower. When
the soft matrix element correction is applied, the distribution shifts
toward smaller $y_{3}$ values since some high $p_{T}$ emissions
will be vetoed by it. Conversely the hard matrix element correction
supplies more events with high $p_{T}$ emissions and so the distribution
shifts toward higher $y_{3}$ values, \emph{}since these events will
not require such fine resolution to distinguish three jets from two
jets. As with the $\Delta R$ distributions, the introduction of the
generalized quasi-collinear splitting functions, $\mathcal{V}_{qq}$,
gives a substantial improvement of the basic parton shower, softening
the jet structure. 

\begin{figure}[t]
\begin{center}\includegraphics[%
  clip,
  width=0.35\textwidth,
  keepaspectratio,
  angle=90]{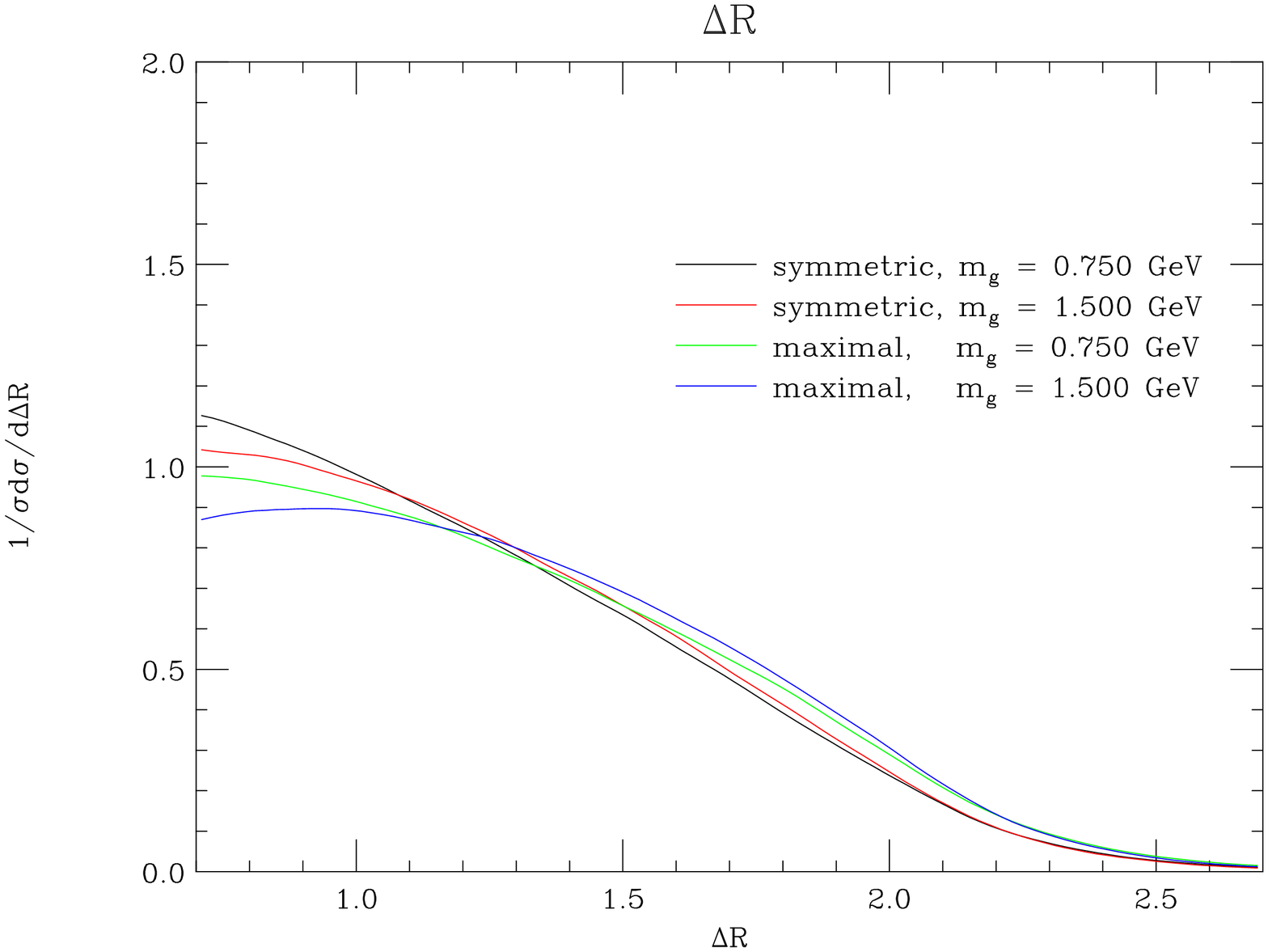}\hfill{}\includegraphics[%
  clip,
  width=0.35\textwidth,
  keepaspectratio,
  angle=90]{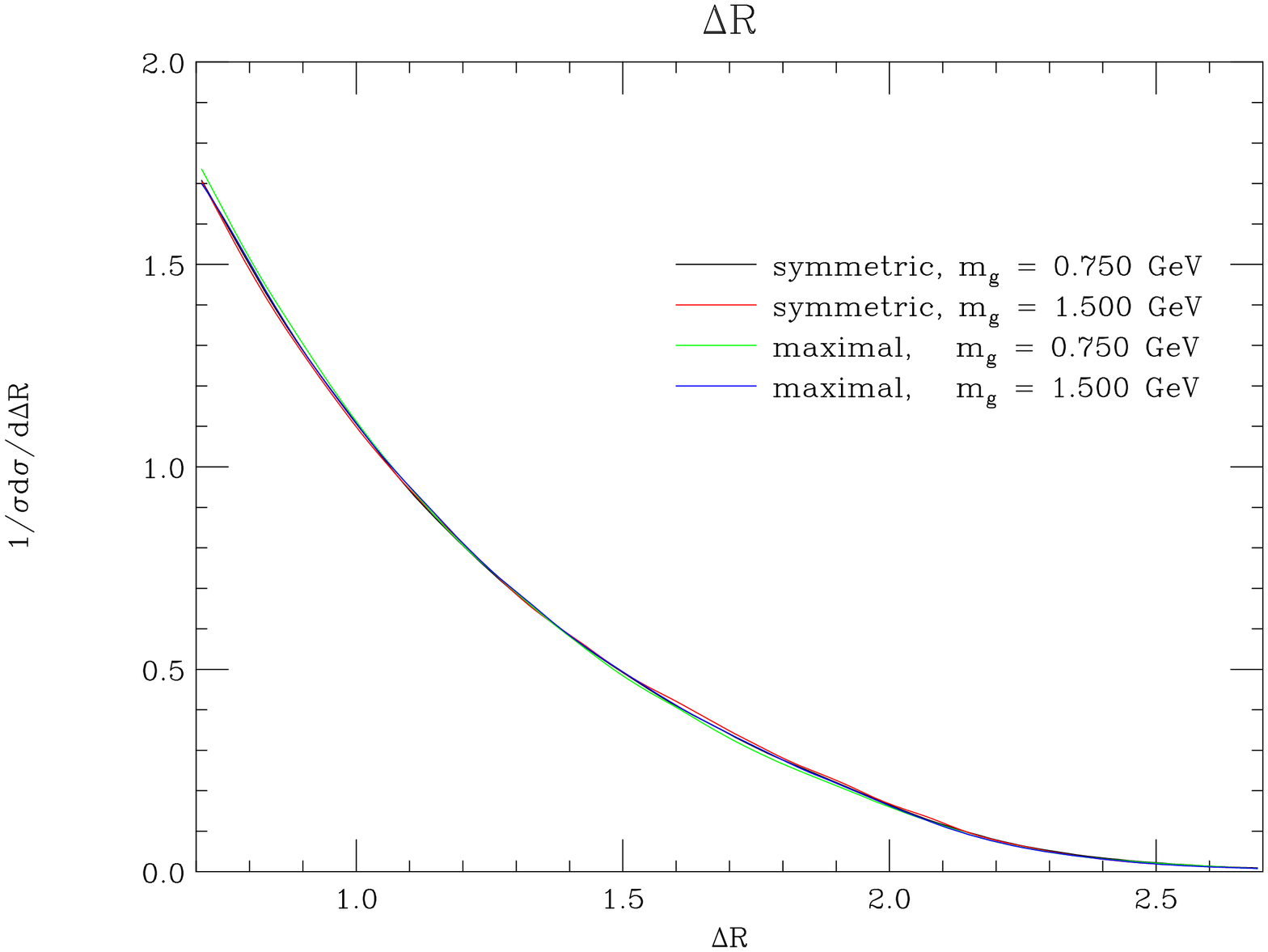}\end{center}

\caption{\label{fig:DeltaR_varying_g}Here we show the $\frac{1}{\sigma}\frac{\mathrm{d}\sigma}{\mathrm{d}\Delta R}$
distribution as in figure \ref{fig:DeltaR}. On the left we show the
distribution obtained from the parton shower approximation and on
the right we show the same distribution including all matrix element
corrections. In each case we have shown how doubling the gluon mass
and/or varying the choice of phase space partition affects the result. }
\end{figure}

\begin{figure}[t]
\begin{center}\includegraphics[%
  clip,
  width=0.35\textwidth,
  keepaspectratio,
  angle=90]{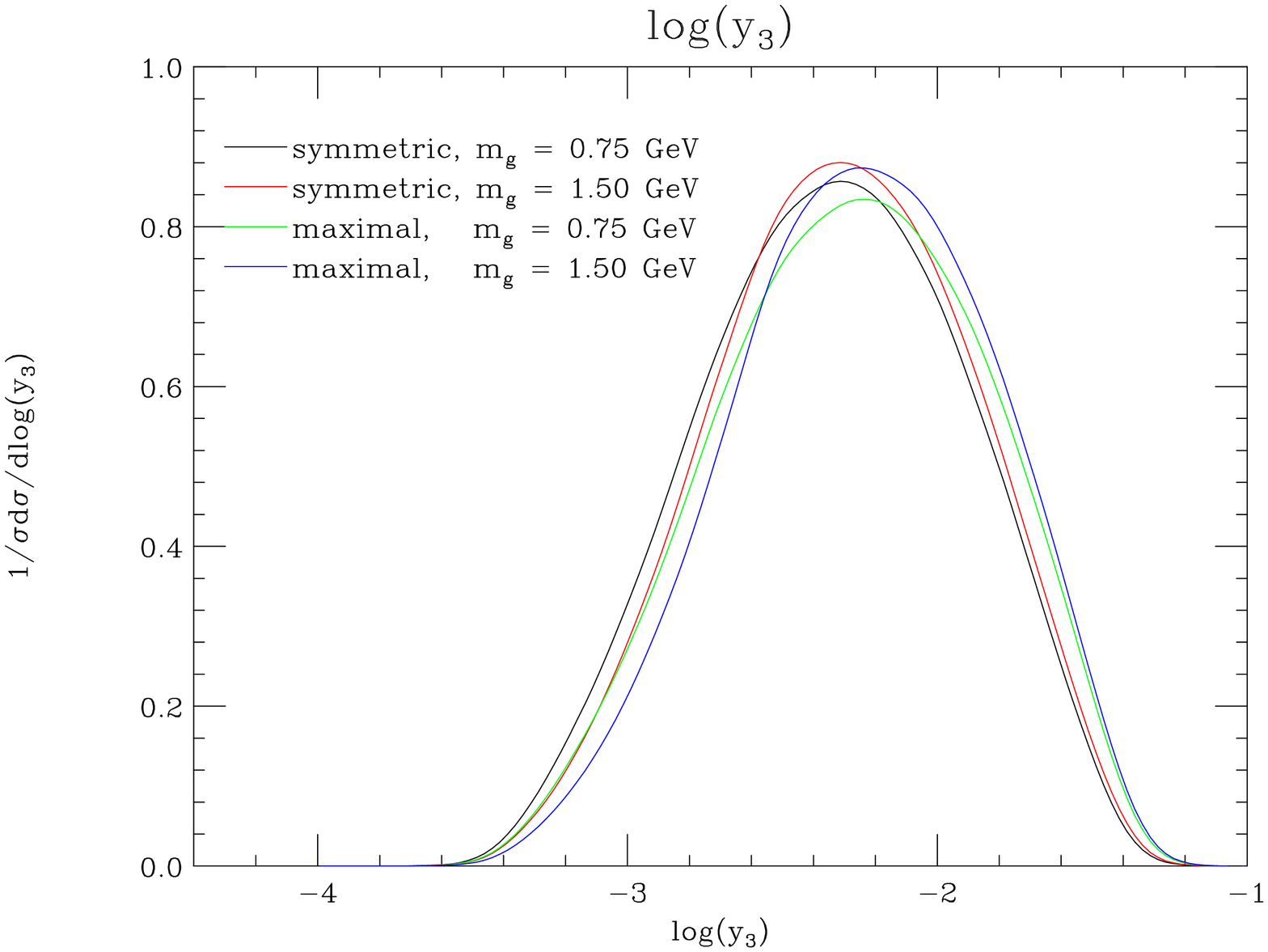}\hfill{}\includegraphics[%
  clip,
  width=0.35\textwidth,
  keepaspectratio,
  angle=90]{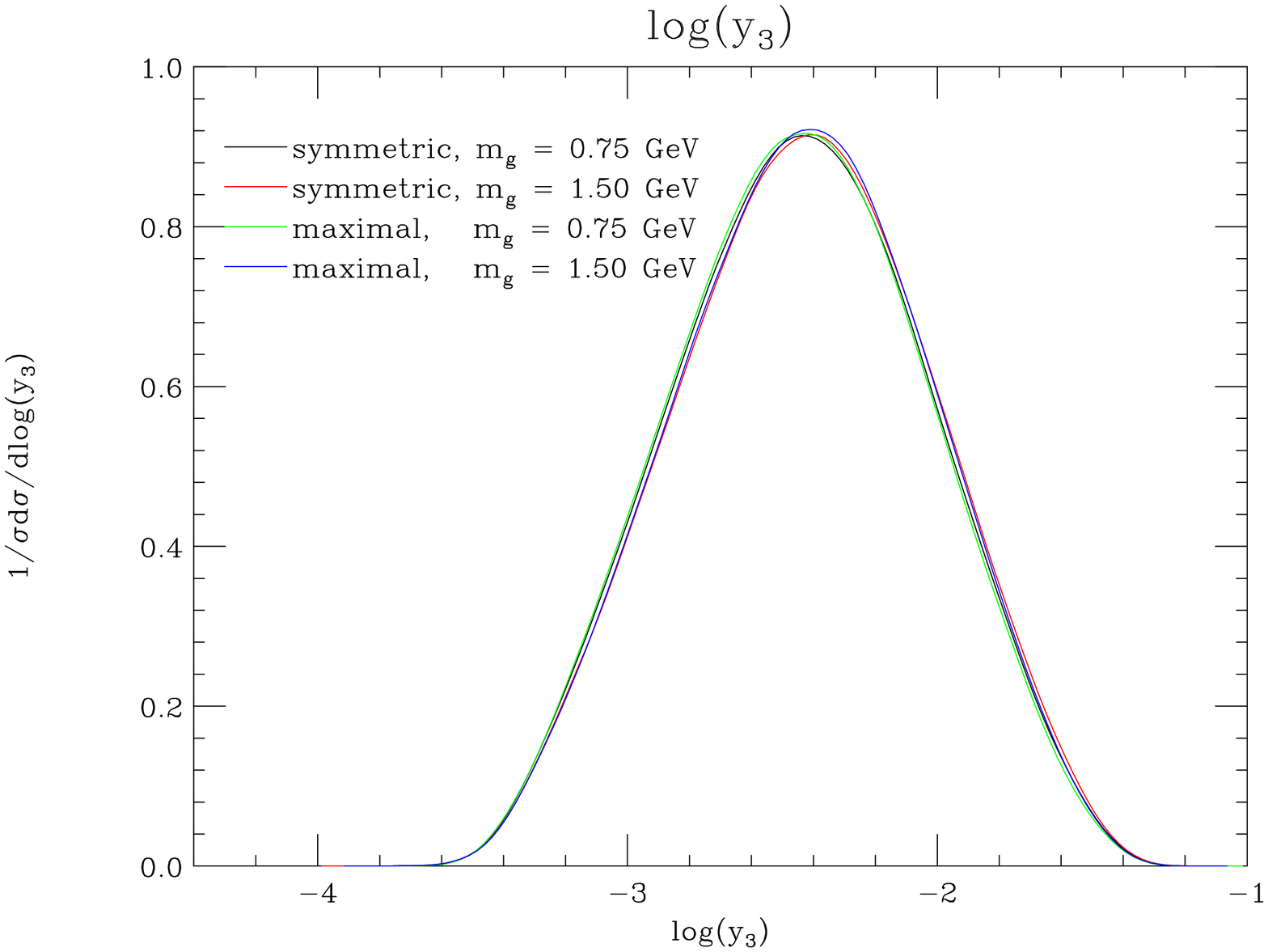}\end{center}

\caption{\label{fig:logy3_varying_g}Here we show the $\frac{1}{\sigma}\frac{\mathrm{d}\sigma}{\mathrm{d}\log\left(y_{3}\right)}$
distribution as in figure \ref{fig:logy3}. On the left we show the
distribution obtained from the parton shower approximation and on
the right we show the same distribution including all matrix element
corrections. In each case we have shown how doubling the gluon mass
and/or varying the choice of phase space partition affects the result. }
\end{figure}

In figures \ref{fig:DeltaR_varying_g} and \ref{fig:logy3_varying_g}
we see the effects of varying the choice of phase space partitioning,
as well as the gluon mass parameter, on the $\Delta R$ and $y_{3}$
distributions. The distributions are less sensitive to changes in
this parameter when the matrix element corrections are applied, compared
to the case of the standalone parton shower. This is to be expected
given that the quasi-collinear splitting functions $\left(P_{qq}\right)$
are known to produce an excess of emissions as one approaches the
acolinear direction. This explains the variation of the distributions
on shifting the phase space boundary, moreover the majority of these
spurious emissions will be soft, resulting in a heightened sensitivity
to the gluon mass parameter.


\section{Conclusions\label{sec:Conclusions}}

At the beginning of this paper we presented a theoretical framework
for the simulation of QCD radiation emitted in top quark decays. In
section \ref{sec:Showering} we described the basic parton shower
formulation. Although this formalism is the product of almost thirty
years of evolution, it has not previously been applied to the extremes
of kinematics and mass scales found in top decay. In doing so we find
that the latest covariant parton shower formalism of \cite{Gieseke:2003rz}
works well under these conditions and we have gained new insight regarding
its use. In particular, we introduced a generalization of the quasi-collinear
limit \cite{Catani:2000ef,Catani:2002hc} for which the $q\rightarrow qg$
splitting function correctly reproduces the eikonal limit.

In section \ref{sec:Matrix-Element-Corrections} we have described
the soft matrix element correction which fixes the distribution of
the \emph{hardest} emission to be that of the leading-order matrix
element for the decay $t\rightarrow bWg$, in a manner respecting
colour coherence. We also describe the hard matrix-element correction
which simply populates the dead region of phase space (not populated
by the parton shower) according to the same single gluon emission
matrix element. 

The results of our program show good agreement with those of the \textsf{FORTRAN}
HERWIG program for the $\Delta R$ and $y_{3}$ observables, provided
that either the soft matrix element correction is applied \cite{Corcella:1998rs},
or our proposed generalization of the $q\rightarrow qg$ quasi-collinear
splitting function is used (\ref{eq:2.4.6}). The associated Dalitz
plots obtained for isolated $t\rightarrow bWg$ decays also meet with
our expectations. Distributions obtained using the standalone parton
shower with the conventional quasi-collinear splitting functions,
suffer from an excess of high $p_{T}$ emissions from the \emph{b-}quark.
This excess in the original basic shower formulation is well understood
to arise from the choice of reference vector $\left(n\right)$ required
for the \emph{b-}quark shower. Future versions of the \textsf{Herwig++}
program will use the new splitting function by default. 

The simulation we have presented improves on that of the older HERWIG
program in a number of ways through the use of the new covariant parton
shower formalism. The new formalism brings with it the use of the
quasi-collinear Altarelli-Parisi splitting functions, which lead to
a natural screening of collinear singularities, allowing us to dispense
with the \emph{ad hoc} angular cut-off, which was responsible for
the spurious \emph{dead}-\emph{cone halo}. Moreover, in this paper
we propose a generalization of these splitting functions, leading
to further improvements in the modelling of soft radiation. We have
also been able to generate radiation from the top quark in its rest
frame, thereby populating infrared regions, corresponding to soft
gluon emissions from the top quark, with the parton shower. Previously
such regions were populated according to a single emission matrix
element correction with an arbitrary soft cut-off \cite{Corcella:1998rs}.

It is clear from our discussion that this work may be readily extended
to decays of other heavy particles, in particular squarks and gluinos.
Should supersymmetry be realised in nature, squark and gluino decays
will give rise to a significant amount of activity in the LHC experiments,
which we will need to simulate. Moreover, as with the top quark today,
measuring the masses of these particles will be a major area of study,
requiring accurate simulations of their decays. It is our intention
to make this extension in a future version of the \textsf{Herwig++}
program.

\section*{Acknowledgements}

We would like to thank our collaborators on the \textsf{Herwig++}
project for many useful discussions and checking our work. This work
was supported by PPARC.

\appendix

\section{Generalizing the $q\rightarrow qg$ quasi-collinear splitting function\label{sec:Generalization_of_Pqq}}

The quasi-collinear limit is that in which $q_{\perp}$ becomes $\mathcal{O}\left(m_{q}\right)$
and small \cite{Catani:2002hc}. This region can be identified by
the \emph{uniform} rescalings $q_{\perp}\rightarrow\lambda q_{\perp}$,
$m_{q}\rightarrow\lambda m_{q}$ and examining the limit $\lambda\rightarrow0$.
In the case of a quasi-collinear gluon emission from a quark, in an
arbitrary $n$ particle process, the squared matrix element factorizes
as\begin{eqnarray}
\lim_{q_{i}\parallel k_{i}}\left|\mathcal{M}_{n}\right|^{2} & = & \frac{8\pi\alpha_{\mathrm{S}}}{q_{i-1}^{2}-m_{q}^{2}}P_{qq}\left|\mathcal{M}_{n-1}\right|^{2},\label{eq:app1}\\
P_{qq} & = & C_{F}\left(\frac{1+z^{2}}{1-z}-\frac{m_{q}^{2}}{q_{i}.k_{i}}\right),\label{eq:app2}\end{eqnarray}
where $z=n.q_{i}/n.q_{i-1}$. 

Note that, as well defining the Sudakov decomposition, the reference
vector $n$ also specifies a choice of axial gauge. Restricting the
form of $n$ imposes restrictions on the choice of axial gauge. To
make calculations easier, $n$ is typically chosen to be lightlike
and perpendicular to $k_{\perp}$, as in \cite{Catani:2000ef}. The
resulting splitting kernel (\ref{eq:app2}) is invariant under gauge
transformations respecting these constraints $\left(n^{2}=n.k_{\perp}=0\right)$
\emph{i.e.} Lorentz boosts of $n$, in the $n$ direction. Invariance
under more general Lorentz transformations requires $k_{\perp}=0$;
in this sense $k_{\perp}$ may be regarded as a measure of the gauge
dependence of approximation (\ref{eq:app1}).

Parton shower simulations use approximation (\ref{eq:app1}) beyond
the collinear limit, so strictly speaking the approach is not gauge
invariant. This is not a problem provided that gauge dependent contributions
are sub-leading. However, in the case of the \emph{b}-quark shower
we have noted an excess of emissions at the edge of the shower phase
space, due to an unphysical singularity, proportional to $n.k_{i}^{-1}$,
where \emph{n} is collinear with the \emph{W} boson. Furthermore,
as discussed in section \ref{sub:Matrix-element-approximations.},
in the limit of soft emissions the splitting kernel (\ref{eq:app1})
does not reduce to the correct eikonal dipole radiation function. 

What we require is that the splitting kernel reproduces the correct
collinear and (ideally) soft limits of the matrix element, without
introducing any other singular terms. Since the behaviour of matrix
elements in these limits is universal, any splitting kernel satisfying
these criteria will be, at least to leading order, gauge invariant. 

It turns out that all of these problems can be solved by simply relaxing
the restrictions on the gauge vector. It is well known that the eikonal
limit of a colour dipole can be calculated by considering gluon emission
from just one quark, provided that the gauge vector is equal to the
momentum of the colour partner. In order to reproduce the correct
soft behaviour we should therefore always set the gauge vector equal
to the four-momentum of the colour partner of the emitter. Although
one can often obtain a good approximation to the eikonal limit by
choosing \emph{n} to be lightlike and collinear to the colour partner,
if the colour partner is heavy, as in top decay, this approximation
fails. 

In any case, this prompts us to consider the case that $n$ is massive.
If we do this we find that the steps leading to (\ref{eq:app1}) now
give, up to sub-leading terms, \begin{eqnarray}
P_{qq}\rightarrow\mathcal{V}_{qq} & = & C_{F}\left(\frac{1+z^{2}}{1-z}-\frac{m_{q}^{2}}{q_{i}.k_{i}}-\frac{n^{2}}{n.k_{i}}\left(\frac{q_{i}.k_{i}}{n.k_{i}}\right)\right),\label{eq:app3}\end{eqnarray}
with $z$ \emph{defined} as $z=n.q_{i}/n.q_{i-1}$. The additional
term in the generalized quasi-collinear splitting function $\left(\mathcal{V}_{qq}\right)$
directly arises from the $n^{2}$ term in the gluon polarization sum.
The new splitting function $\mathcal{V}_{qq}$ (\ref{eq:app3}) reproduces
the correct soft and collinear limits in the shower phase space, it
contains no unphysical divergent terms. 

From a purely pragmatic point of view $\mathcal{V}_{qq}$ can be considered
as a kind of global soft matrix element correction and it is seen
to have similar effects on physical observables (see section \ref{sec:Results}).
Crucially, unlike the soft matrix element corrections, the generalized
quasi-collinear splitting function is process-independent.

\section{\emph{t}$\rightarrow$\emph{bWg} phase-space and matrix element\label{sec:t-bWg-phase-space}}

In this appendix we give the matrix element for $t\rightarrow bWg$
decay and the phase-space parametrization, these are necessary for
the hard matrix element correction discussed in section \ref{sec:Matrix-Element-Corrections}.
Both the phase-space and matrix element are parametrized in terms
of the Dalitz variables \begin{equation}
x_{i}=\frac{2q_{i}.p_{t}}{m_{t}^{2}},\label{eq:6.1}\end{equation}
which, in the top quark rest frame, are equal to two times the fraction
of the top quark's energy carried by particle $i$. We also define
the following ratios of masses for convenience:\begin{equation}
\begin{array}{lclclcl}
b & = & \frac{m_{b}^{2}}{m_{t}^{2}} & , & w & = & \frac{m_{W}^{2}}{m_{t}^{2}}.\end{array}\label{eq:6.2}\end{equation}

\subsection{Phase space\label{sub:Phase-space} }

The Dalitz variables $x_{i}$, were calculated in \cite{Gieseke:2003rz},
in terms of $z$ and the top quark evolution variable $\tilde{\kappa}_{t}=\tilde{q}^{2}/m_{t}^{2}$,
assuming the gluon was emitted by the top quark, as being

\begin{equation}
\begin{array}{rcl}
x_{W}\left(z,\tilde{\kappa}_{t}\right) & = & \frac{1+w-b-\left(1-z\right)\tilde{\kappa}_{t}-\sqrt{\left(1+w-b-\left(1-z\right)\tilde{\kappa}_{t}\right)^{2}-4w\left(1-\left(1-z\right)\left(\tilde{\kappa}_{t}-1\right)\right)z}}{2z}\\
 & + & \frac{1+w-b-\left(1-z\right)\tilde{\kappa}_{t}+\sqrt{\left(1+w-b-\left(1-z\right)\tilde{\kappa}_{t}\right)^{2}-4w\left(1-\left(1-z\right)\left(\tilde{\kappa}_{t}-1\right)\right)z}}{2\left(1-\left(1-z\right)\left(\tilde{\kappa}_{t}-1\right)\right)}\\
x_{g}\left(z,\tilde{\kappa}_{t}\right) & = & \left(1-z\right)\tilde{\kappa}_{t},\end{array},\label{eq:6.1.1}\end{equation}
where $b=m_{b}^{2}/m_{t}^{2}$ and $w=m_{W}^{2}/m_{t}^{2}$. We may
completely eliminate $z$ from $x_{W}$ to give \begin{equation}
\begin{array}{rcl}
x_{W}\left(x_{g},\tilde{\kappa}_{t}\right) & = & \frac{1}{2\left(\tilde{\kappa}_{t}-x_{g}\right)}\left(\tilde{\kappa}_{t}\left(1+w-b-x_{g}\right)-\Lambda\left(x_{g},\tilde{\kappa}_{t}\right)\right)\\
 & + & \frac{1}{2\left(\tilde{\kappa}_{t}+x_{g}\left(1-\tilde{\kappa}_{t}\right)\right)}\left(\tilde{\kappa}_{t}\left(1+w-b-x_{g}\right)+\Lambda\left(x_{g},\tilde{\kappa}_{t}\right)\right),\\
\Lambda\left(x_{g},\tilde{\kappa}_{t}\right) & = & \sqrt{\left(x_{g}-\hat{x}_{g_{+}}\right)\left(x_{g}-\hat{x}_{g_{-}}\right)\left(\tilde{\kappa}_{t}-\tilde{\kappa}_{+}\right)\left(\tilde{\kappa}_{t}-\tilde{\kappa}_{-}\right)},\\
\hat{x}_{g_{\pm}} & = & 1-\left(\sqrt{w}\pm\sqrt{b}\right)^{2},\\
\tilde{\kappa}_{\pm} & = & 2x_{g}\left(x_{g}\pm\sqrt{\left(1-\frac{1}{w}\right)\left(x_{g}-\bar{x}_{g_{+}}\right)\left(x_{g}-\bar{x}_{g_{-}}\right)}\right)^{-1},\\
\bar{x}_{g_{\pm}} & = & \frac{\left(1-w\right)\left(1\pm\sqrt{w}\right)-b\left(1\mp\sqrt{w}\right)}{1-w}.\end{array}\label{eq:6.1.2}\end{equation}
The expression for $x_{W}$ given in (\ref{eq:6.1.2}) enables one
to draw lines of constant $\tilde{\kappa}_{t}$ in the $x_{W}$, $x_{g}$
plane. 

Repeating the procedure for the case that the gluon is assumed to
originate from the \emph{b}-quark gives\begin{equation}
\begin{array}{lcl}
x_{g_{\pm}}\left(x_{W},\tilde{\kappa}_{b}\right) & = & 2-x_{W}-z_{\pm}\left(x_{W},\tilde{\kappa}_{b}\right)\sqrt{x_{W}^{2}-4w}\\
 & - & \frac{1}{2}\left(1+\frac{b}{1+w-x_{W}}\right)\left(2-x_{W}-\sqrt{x_{W}^{2}-4w}\right),\\
z_{\pm}\left(x_{W},\tilde{\kappa}_{b}\right) & = & \frac{1}{2\tilde{\kappa}_{b}}\left(\tilde{\kappa}_{b}\pm\sqrt{\tilde{\kappa}_{b}^{2}-4\tilde{\kappa}_{b}\left(1+w-b-x_{W}\right)}\right).\end{array}\label{eq:6.1.3}\end{equation}
Inverting (\ref{eq:6.1.3}) to obtain $x_{W}$ as a function of $x_{g}$
involves a high-order polynomial requiring a numerical solution, neglecting
the \emph{b}-quark mass an analytic solution becomes possible.

\subsection{Matrix element\label{sub:Matrix-element.}}

The matrix element for the decay $t\rightarrow bWg$ was given in
\cite{Corcella:1998rs,Gieseke:2003rz} assuming a massless \emph{b}-quark.
We calculate the squared matrix element, without neglecting the \emph{b}-quark
mass and find\begin{equation}
\begin{array}{rcl}
\frac{1}{\Gamma_{0}}\frac{\mathrm{d}^{2}\Gamma}{\mathrm{d}x_{g}\mathrm{d}x_{W}} & = & \frac{\alpha_{S}C_{F}}{\pi}\frac{1}{\lambda\bar{x}_{W}x_{g}^{2}}\\
 &  & \left(-\frac{bx_{g}^{2}}{\bar{x}_{W}}+\left(1-w+b\right)x_{g}-\bar{x}_{W}\left(1-x_{g}\right)-x_{g}^{2}\right.\\
 &  & \left.+\frac{x_{g}}{1+w-2w^{2}-b\left(2-w-b\right)}\left(\frac{1}{2}\left(1+2w+b\right)\left(\bar{x}_{W}-x_{g}\right)^{2}+2w\bar{x}_{W}x_{g}\right)\right)\\
\bar{x}_{W} & = & 1+w-b-x_{W}\end{array},\label{eq:6.2.1}\end{equation}
where again, $b=m_{b}^{2}/m_{t}^{2}$ and $w=m_{W}^{2}/m_{t}^{2}$.
Setting $b=0$ , our expression (\ref{eq:6.2.1}) easily reduces to
those given in \cite{Corcella:1998rs,Gieseke:2003rz}.

\bibliographystyle{JHEP}
\bibliography{TopPaper}

\end{document}